\documentclass[preview,12pt]{elsarticle}

\usepackage{graphicx}
\usepackage{amssymb}

\usepackage{times}
\usepackage{amsmath}
\usepackage{color}

\usepackage{empheq}
\usepackage{bm}

\usepackage{color}
\usepackage{tikz}

\usepackage{lineno}
\usepackage[normalem]{ulem}

\usetikzlibrary{shapes.geometric, arrows}
\makeatletter
\def\els@aparagraph[#1]#2{\elsparagraph[#1]{#2}} 
\def\els@bparagraph#1{\elsparagraph*{#1}}        
\makeatother

\newcommand{\code}[1]{{\bf #1}}
\newcommand{\figref}[1]{Fig.~\ref{#1}}
\newcommand{\figsref}[2]{Figs.~\ref{#1}--\ref{#2}}

\newcommand{\newtext}[1]{{#1}}
\newcommand{\oldtext}[1]{}

\journal{Computers \& Structures}

\begin{document}


\begin{frontmatter}

\title{Inverse analysis of traction-separation relationship based on sequentially linear approach}

\author{Jan Vorel}
\address{Associate Professor, Department of Mechanics, Faculty of Civil Engineering, Czech Technical University in~Prague, Th\'{a}kurova 7, 166\,29 Praha 6, Czech Republic. Email:~jan.vorel@fsv.cvut.cz (corresponding author)}

\author{Petr Kabele}
\address{Professor, Department of Mechanics, Faculty of Civil Engineering, Czech Technical University in~Prague, Th\'{a}kurova 7, 166\,29 Praha 6, Czech Republic. Email:~petr.kabele@fsv.cvut.cz}

\begin{abstract}
  Traction-separation relationship is an important material characteristic describing the fracture behaviour of quasi-brittle solids. A new numerical scheme for identification of the traction-separation relation by inverse analysis of data obtained from various types of fracture tests is proposed.  Due to employing the concept of sequentially linear analysis, the method exhibits a superior numerical stability and versatility. The applicability and effectiveness of the proposed method is demonstrated on examples involving identification of the traction-separation relationship using experimental data from various test configurations.
\end{abstract}


\end{frontmatter}

\section{Introduction}
\label{sec:intro}
\begin{figure}
\begin{center}
\begin{tabular}{cc}
  \includegraphics[width=0.4\textwidth]{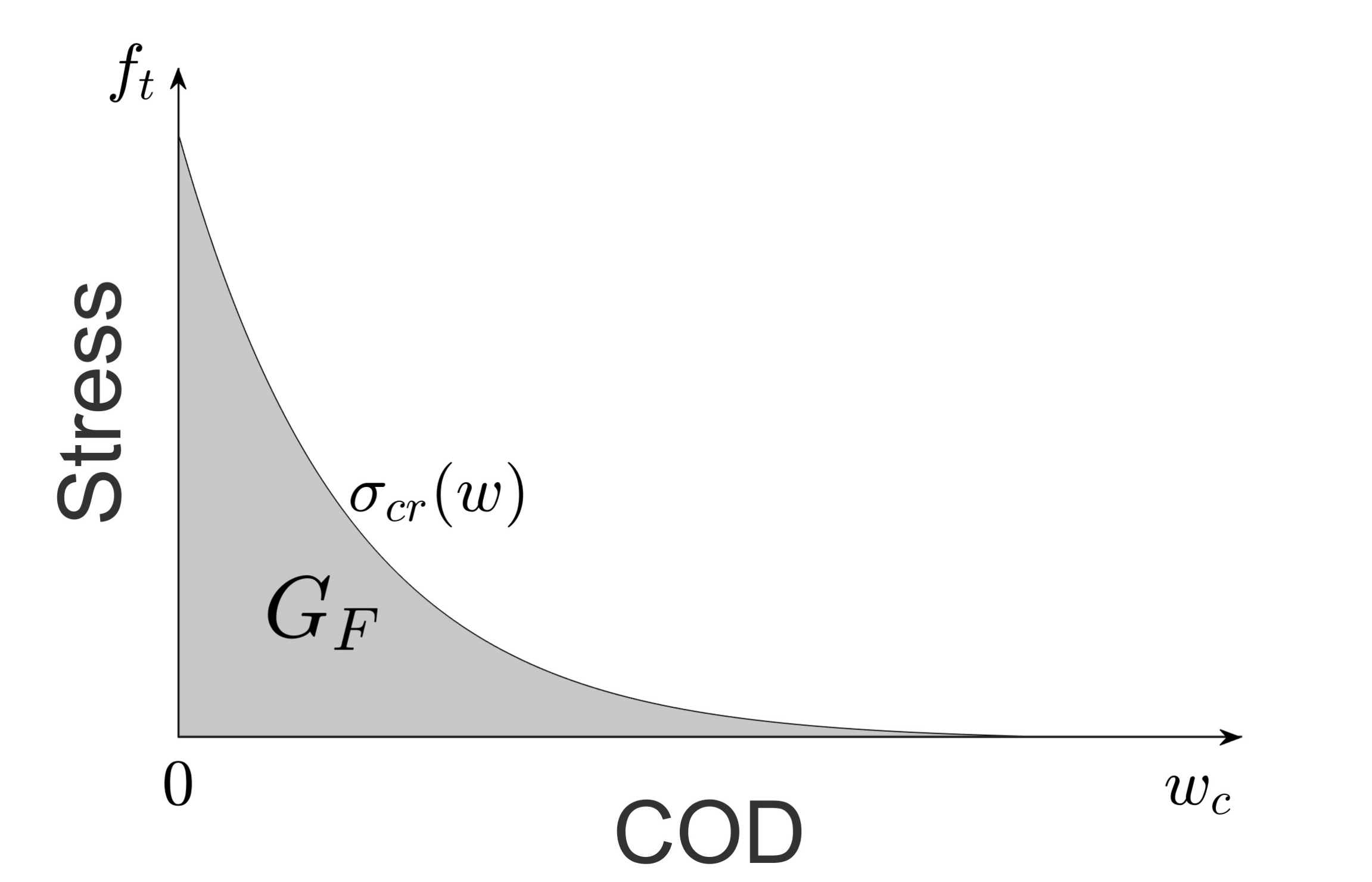}
\end{tabular}
\caption{Stress-COD diagram}
\label{fig:Hillerborg}
\end{center}
\end{figure}

Since it was proposed by~\citet{Hillerborg:1976:analysis}, the cohesive crack model has gained a great prominence in numerical analysis and simulation of fracture in quasi-brittle materials, such as, concrete, fibre-reinforced concrete, rock, masonry, wood, etc. The cohesive crack is envisioned as a discontinuity in a displacement field, across which a cohesive stress is transferred. Physically, this transfer is attributed to the bridging action of ligaments, such as aggregate particles, fibres, contacts of tortuous crack surfaces, etc. Mathematically, this effect is represented by the, so-called, traction-separation (TS) relationship (or law, also termed cohesive law or bridging law). For mode-I fracture, the law relates the crack-normal component of the cohesive stress to the crack opening displacement (COD): $\sigma_{cr}(w)$ (\figref{fig:Hillerborg}). The TS relationship is thus the key material characteristic of the model. The first point of the TS relation corresponds to the tensile strength $f_{t}=\sigma_{cr}(0)$. The area enclosed under the TS diagram represents the fracture energy:
\begin{equation}
G_{F}=\int_{w=0}^{w_{c}}\sigma_{cr}(w)dw,
\end{equation}
where $w_{c}$ is the COD at which the bridging effect is exhausted and the crack becomes free of traction.

The importance of determining the TS relationship from experiments has been recognised over years by many researchers. A comprehensive summary and evaluation of existing methods for concrete was carried out by RILEM Technical Committee 187-SOC, which was launched in 2001. The final report \cite{Planas:2007:report39} deals both with aspects of experimental testing as well as with subsequent evaluation of TS relation either directly or by means of inverse analysis (IA).

The direct tension test, in which \newtext{a} crack is induced on a uniformly stressed cross-section of an elongated specimen, appears as the most natural way to determine the TS relation experimentally. Nevertheless, its execution poses several difficulties. First of all, the test must capture the softening response after cracking up to complete separation. This implies that a stiff loading frame and displacement control must be employed. Even under these conditions it may be difficult to obtain stable crack opening if the TS relation exhibits a steep softening and the specimen has accumulated a large elastic energy before cracking, which results in snap-back in the load-displacement relationship. It is also tricky to attach the specimen to the loading machine in such a way that there are no bending and twisting effects, both before and during cracking. These difficulties are further emphasised due to inherent material inhomogeneity, which causes non-uniform cracking.

For the above-cited reasons it is often preferable to test fracture in quasi-brittle materials on other, more feasible, configurations, such as three- or four-point bending of notched beam, wedge splitting or compact tension. These configurations, however, do not allow direct evaluation of the TS relationship, as the crack opening displacement and distribution of the cohesive stress are not uniform during the test. Usually, overall response in terms of load vs. deflection or load vs. crack mouth opening displacement (CMOD) is recorded. Therefore, several methods, different in complexity and experimental data requirements, have been developed to retrieve the TS relation from tests involving propagation of cracks under non-uniform stress conditions.

The technique introduced by~\citet{Li:1989:NT} utilises the mathematical relation between the TS law and the J-integral evaluated along a path enclosing the cohesive zone. First, the relationship between the J-integral and the notch tip opening displacement $J(w)$ is obtained from measurements on two fracture specimens with sightly different initial notch size. To this end, the corresponding values of load, load point displacement and notch tip separation must be simultaneously recorded. Secondly, the $J(w)$ relation is numerically differentiated, which results in the desired $\sigma_{cr}(w)$ relation.

Another methodology is based on assuming a certain functional form of the TS relation and obtaining particular values of its parameters by fitting the global response (load vs.~deflection or load vs.~CMOD) of a model to the experimentally obtained one. The most commonly used forms of the TS relation are linear, piece-wise linear, or Hordijk's exponential \cite{Hordijk:1991:local}. While modelling the experiment, it is usually assumed that the fracture occurs on a single plane while the material outside the cracking zone is elastic. To this end, various approaches have been used, including the cracked hinge model, e.g.~\cite{Ostergaard:2003:earlyage,Skocek:2008:Inverse,Sousa:2006:determining,Jepsen:2016:afully} or the finite element method (FEM), e.g \cite{Slowik:2006:IA}. The fitting is achieved by minimising an objective function using various optimisation techniques, such as the gradient methods with a line search \cite{Sousa:2006:determining}, the simplex method \cite{Skocek:2008:Inverse} or the evolutionary algorithm including local neighbourhood attraction \cite{Slowik:2006:IA}.

Yet another approach to \newtext{determine} \oldtext{determining} the TS relationship was proposed by~\citet{Nanakorn:1996:Back}. In this case, the TS law is approximated by a continuous piece-wise linear function. A finite element (FE) model employing the so-called ``cracked element'' is used to incrementally simulate a fracture test with a monotonously propagating and opening crack. In each increment, the slope of the TS relation in the element just ahead of the traction-free notch is determined so that, the slopes of the corresponding segments of the load-deflection curves from the simulation and from the experiment match. As the COD of the element at the notch tip monotonously increases from zero, individual segments of the TS curve are identified. Furthermore, because the elements further along the crack path exhibit smaller COD than the notch-tip element, they can be assigned the portion of the TS relation already identified from the first element. It should be noted that almost simultaneously, but independently, a similar method was also developed by~\citet{Kitsutaka:1995:fracture}. This method was eventually adopted by the Japan Concrete Institute~\cite{JCI:2003:japanconcrete}.

Each of the discussed methodologies has its own advantages and disadvantages. The J-integral method \cite{Li:1989:NT} is computationally least demanding, as it involves only numerical differentiation of the experimental data and the experiments need not to be numerically simulated. On the other hand, it requires that two tests with different notch sizes are performed for each evaluation of TS relation. Thus, results may be affected by the choice of the notch size difference as well as by the inevitable material variability of the two tested specimens.

\newtext{As the global optimization based methods~\cite{Ostergaard:2003:earlyage,Skocek:2008:Inverse,Sousa:2006:determining,Jepsen:2016:afully,Slowik:2006:IA} aim for matching the general trend of the measured load displacement relation, they are rather insensitive to its local irregularities resulting from local inhomogeneities along the crack path. Thus, the methods inherently "smooth out" the TS relationship, which then represents the overall cohesive property of the ligament material.} \oldtext{The main feature of the global optimisation based methods [5,6,7,8] is that they utilise the complete measured load-displacement curve, which usually represents propagation of the crack through the entire section of the tested specimen. The method thus inherently ``smooths out'' any local variations of the TS relation which might result from material inhomogeneity.} However, the necessity to assume a specific form of the TS relation and to constrain the number of free parameters for the optimisation to be feasible could be seen as disadvantages of these methods. It is also not guaranteed that the procedure identifies the global minimum of the multidimensional objective function, which means that non-unique solutions may be found (e.g. depending on the choice of initial values of parameters).

The incremental inverse analysis \cite{Nanakorn:1996:Back,Kitsutaka:1995:fracture} provides a unique solution for a given discretization of the experimental data, as the slope of the TS relation is related to the slope of the load-deflection curve by a closed-form expression~\cite{Nanakorn:1996:Back}. The main disadvantage of this method consists in the fact that the TS relation is identified based on the local cohesive behaviour of a small part of the fracture surface just ahead of the notch (the notch-tip element). This relation is then applied to the entire ligament, which, however, in the experimental specimen is not perfectly homogeneous. As a consequence, the obtained TS relations exhibit severe oscillations and need to be smoothed~\cite{Nanakorn:1996:Back}. Another implication is that only the initial part of the load-displacement data, until the notch-tip element becomes stress free (its COD reaches the value of $w_{c}$), is used for the identification of the TS law.

It should be noted that many of the earlier-discussed inverse methods involve finite element solution of a nonlinear problem. A sequentially linear analysis (SLA), also called ``event-by-event'' scheme, has been recently developed to overcome some numerical difficulties arising when the finite element method is used to model structures involving quasi-brittle materials. This scheme is particularly well suited to large-scale structural analyses~\cite{Rots:2004:RSLSM}. Simulating fracture through the event-by-event cracking procedure is an attractive alternative to standard nonlinear solution algorithms, such as the Newton-Raphson or arc-length method. The calculation proceeds in the form of a sequence of linear steps, while an increment of fracture, as opposed to increment of force or displacement, is imposed on the model in each step. Therefore, large jumps in cracking during a single load/time step, which can be a source of convergence problems and can significantly influence results, can be avoided. In other words, while the conventional nonlinear finite element analysis ‘‘skips over” portions of the structural response when brittle behaviour occurs and rejoins the response through iteration algorithms, event-by-event procedure eschews this by controlling damage evolution directly~\cite{Vorel:2015:CM,Vorel:2014:SHCC}. \newtext{This is achieved by imposing, in each step, incremental extension of crack path or increased of crack opening at the most critical location of the model (typically an integration point) and performing linear analysis with secant material stiffness. Thus, a nonlinear TS law is replaced by its "saw tooth" representation. By performing multiple steps of this procedure, a nonlinear (possibly softening) overall structural response, is obtained.}

In this paper we propose a new numerical scheme for the determination of TS relationship by an inverse analysis. The proposed methodology is, in principle, based on the idea of~\cite{Nanakorn:1996:Back,Kitsutaka:1995:fracture}, but it is implemented in the framework of SLA. In contrast to the previous works, the method brings the following advantages:
\begin{itemize}
  \item It includes calculation of tensile strength, i.e. it does not require that this parameter is pre-defined beforehand.
  \item Any material model compatible with SLA can be used to represent the behaviour outside the fracture zone \newtext{(not only elastic constitutive models, such as isotropic or orthotropic, but also inelastic ones, such as plasticity, damage, etc.)}
  \item It accepts any kind of \newtext{proportional} loading vs. response data pair uniquely characterising the loading path (e.g. load vs. deflection, load vs. CMOD, load-point displacement vs. CMOD) and thus offers higher versatility in terms of test control, experimental configurations and measured experimental data.
  \item As only linear solution is needed for each calculation step \newtext{(see following sections)}, the method can be either easily implemented into or used with a variety of FE softwares and exhibits better numerical stability.
\end{itemize}

An additional improvement consists in applying the inverse analysis in multiple passes, which makes it possible to systematically utilise the entire measured load-displacement curve, even beyond the point when the notch-tip element becomes stress-free, to improve accuracy of the solution.

The applicability and effectiveness of the proposed method is demonstrated on examples involving identification of the TS relationship from experimental data of various test configurations (three-point bending and compact tension) and different materials (fibre reinforced cement-based composites, wood).

\section{Principle of sequentially linear analysis}
\label{sec:sla}
\begin{figure}
\begin{center}
\begin{tabular}{cc}
  \includegraphics[width=0.4\textwidth]{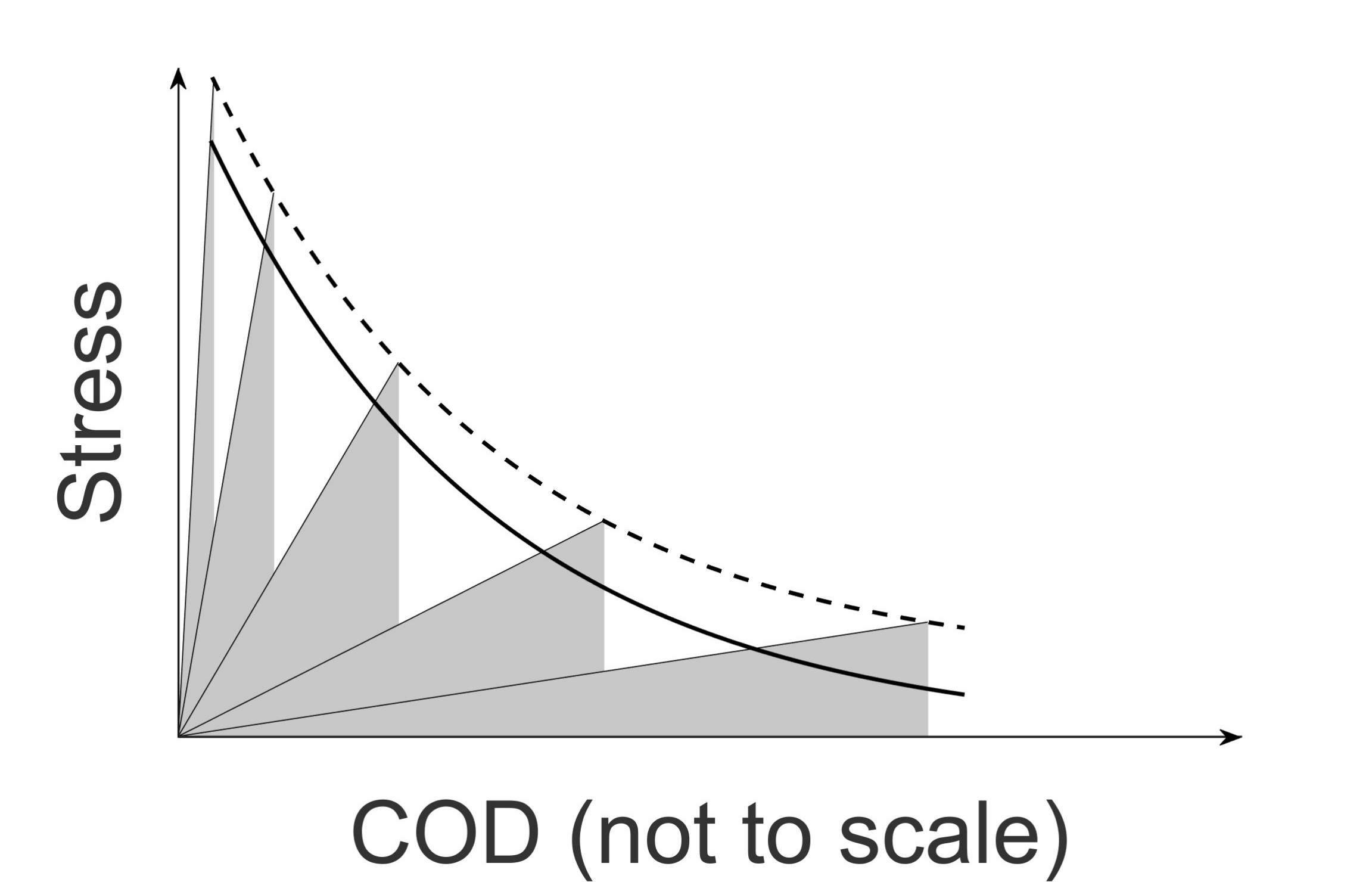} &
  \includegraphics[width=0.4\textwidth]{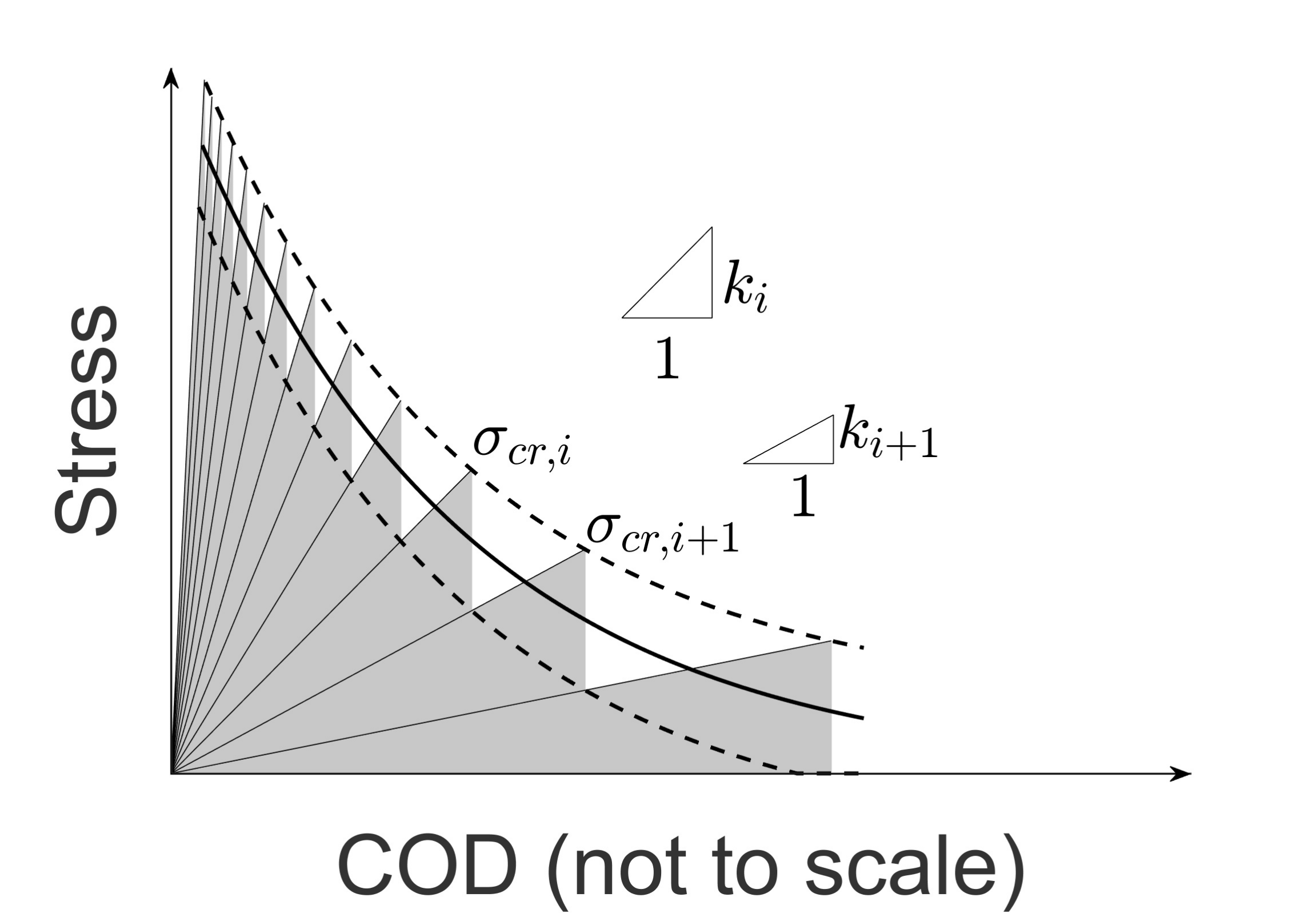} \\
  (a) & (b)
\end{tabular}
\caption{``Saw-tooth'' approximation of softening (initial penalty stiffness is reduced for clarity): (a)~reduction of normal stiffness by a fixed factor; (b)~stress-band model}
\label{fig:SLA}
\end{center}
\end{figure}

\newtext{In this section we review the principle of the ``direct'' sequentially linear analysis. This sets the ground for its utilization in the framework of inverse analysis, which will be discussed in the subsequent sections. The sequentially linear analysis was specifically developed to overcome numerical difficulties, arising from a steep softening or even a snap-back behaviour, which are often encountered when quasi-brittle structures are analysed by FEM~\cite{Rots:2004:RSLSM,Rots:2004:SAW}. The method is generally applicable to quasi-brittle materials with a linear or nonlinear softening, jumps in a cohesive law, or even to strain-hardening brittle-matrix composites, for all of which the method is capable of tracing the progressive failure process in a robust manner~\cite{Dejong:2008:SLA,Billington:2009:ECC}. It can be used in combination with various nonlinear material models, including continuum damage, incremental plasticity, or cohesive crack. Hereafter, we focus on implementation of SLA with the cohesive crack model, which is characterized by tensile strength and TS relationship.

The SLA procedure is based on the solution of physically nonlinear problems by a sequence of linear (secant) steps. For this purpose, a given TS relation is approximated by a so-called “saw-tooth” law. The saw-tooth law is a collection of elastic-brittle teeth, which maintain a positive tangent stiffness. There are different ways in which the saw-tooth law can be derived from a TS diagram. For example, the TS diagram can be imitated by consecutively reducing stiffness (\figref{fig:SLA}(a)), or by defining a stress band along the softening curve (\figref{fig:SLA}(b)); see~\cite{Invernizzi:2011:SLM} for more details. The stepping is executed by so-called event-by-event procedure. In each step, the cracking progresses in only one, the most critical, integration point of the FE-discretized model.}


%
%


\newtext{
The SLA procedure can be summarised as follows~\cite{Rots:2004:RSLSM,Graaf:2017}:
\begin{enumerate}
  \item A saw-tooth law (e.g. \figref{fig:SLA}) representing a given TS relationship is defined and assigned to integration points of all elements, in which cracking may potentially occur. These may be all elements in the model, if the crack path is not a-priori constrained, or only those elements, lying along anticipated crack path.
  \item A reference load is applied on the structural FE model and corresponding deformations and stresses are calculated at each of the above-mentioned integration points through a linear-elastic analysis.
  \item The critical integration point and the corresponding critical load multiplier are determined as follows. The calculated stress in each integration point is compared with its actual strength (fracture condition). This may be the initial strength for a point that has not cracked yet, or the residual strength defined by the saw-tooth law for a point that is already cracked. The critical integration point is identified as the one, in which the stress has to be scaled by the smallest factor to satisfy the fracture condition. The value of this factor then corresponds to the critical load multiplier $\lambda$ for the actual step. Note that if multiple integration points have the same lowest value of $\lambda$, only one integration point is randomly chosen as the critical one.
  \item The reference load is proportionally scaled by the critical load multiplier ($\lambda$) and the current stress and strain state is determined. At this state, the fracture condition is satisfied only at the critical point, while in all other integration points of the model no further cracking occurs.
  \item An increment of cracking is applied at the critical integration point by updating its state according to the saw-tooth law, see~\figref{fig:SLA}(b). Its secant stiffness is reduced $\left(k_{cr,i}\rightarrow k_{cr,i+1}\right)$ and the new residual strength $\left(\sigma_{cr,i}\rightarrow\sigma_{cr,i+1}\right)$ is assigned to it.
  \item The previous sequence of steps (2-5), in which properties of a single (critical) integration point are updated in each cycle, is repeated until the final prescribed load or displacement of the structure is reached. The nonlinear response of the structure is obtained by linking consecutively the results of all cycles.
\end{enumerate}
}

\newtext{The major drawback of ``standard'' SLA is the inability to properly capture the non-proportional loading. An extension of the SLA concept towards the non-proportional loading was proposed in~\cite{Dejong:2008:SLA,Giardina:2013:NAM,Elias:2010:ISLA,Elias:2015:GLUFR}. However, these extensions are not discussed in the present paper since only the experimental results obtained under proportional loading will be used for the inverse analysis.}

\section{Inverse analysis}
\label{sec:backAnal}

\subsection{Principle of method}
\label{subsec:principle}

As already mentioned, the proposed methodology of inverse solution of TS relation is based on the sequentially linear analysis, see Section~\ref{sec:sla} and the incremental approach of~\citet{Nanakorn:1996:Back}, see Section~\ref{sec:intro}.
As opposed to~\citet{Nanakorn:1996:Back}, who employed a ``cracked element'' with displacement discontinuity embedded in its shape functions, we use zero-thickness cohesive interface elements~\cite{PatBit01} to model cracks. Interface elements are introduced along potential crack paths in a finite element model and the corresponding TS relation for these elements is determined during the analysis. Note that the interface elements are initially assigned penalty stiffnesses (both normal $k_0$ and shear $g_0$) to represent undamaged material. This initial penalty stiffnesses should be set sufficiently high to limit the interface opening or overlap and sliding to negligible levels. At the same time, their values should not be excessively high to avoid ill-conditioning of the global stiffness matrix. The identified TS relation thus includes the separation, presumably very small, that the cohesive element acquires before its cracking starts. Elements outside the potential crack paths may be linear elastic, or they can even represent nonlinear material behaviour, provided the corresponding “saw-tooth” approximation is known. \newtext{In general, the properties of bulk material are prescribed by the user. For the numerical examples presented in Section~\ref{sec:numStudy}, the value of Young's modulus is always determined by matching the numerical response to the initial elastic part of input experimental data. Note that the elastic numerical response has to overlap or be below the experimental data, i.e. the assigned Young modulus is less or equal to the true one, otherwise, as mentioned later, the criteria for obtaining TS diagram would not be fulfilled.}

The proposed procedure utilises as an input experimentally acquired values of any two variables uniquely characterising the evolution of the applied loading and the corresponding response of the structure. These can be, e.g. prescribed force load vs. measured displacement; prescribed force load vs. measured CMOD; prescribed displacement vs. measured displacement; etc. In the subsequent text we will refer to the prescribed variable as “load” and the measured variable as “displacement”. During the analysis, values of these variables calculated by the finite element model are matched to the experimental data, by which the TS relation is determined. As the experimental data are always in the form of discrete values, a linear interpolation is used whenever intermediate values are sought.
In the subsequent paragraphs, we describe the procedure of inverse analysis assuming that the TS relation for mode I, i.e. crack-normal stress $\sigma_{cr}$ vs. \oldtext{normal displacement jump} \newtext{COD} $w$ \oldtext{(COD)} is being identified. To this end, the experiment should be configured in such a way that the crack propagates and opens in the pure mode I. Identification of mode II relation (shear stress vs. crack slip) would follow, in principle, the same steps. However, the experiment would have to capture pure mode II cracking under the constant crack-normal stress. The identification of a fully coupled normal-shear TS relation is beyond the scope of this paper. The calculation is performed in the following steps.

\begin{figure}
\begin{center}
\begin{tabular}{ccc}
  \includegraphics[scale=0.45]{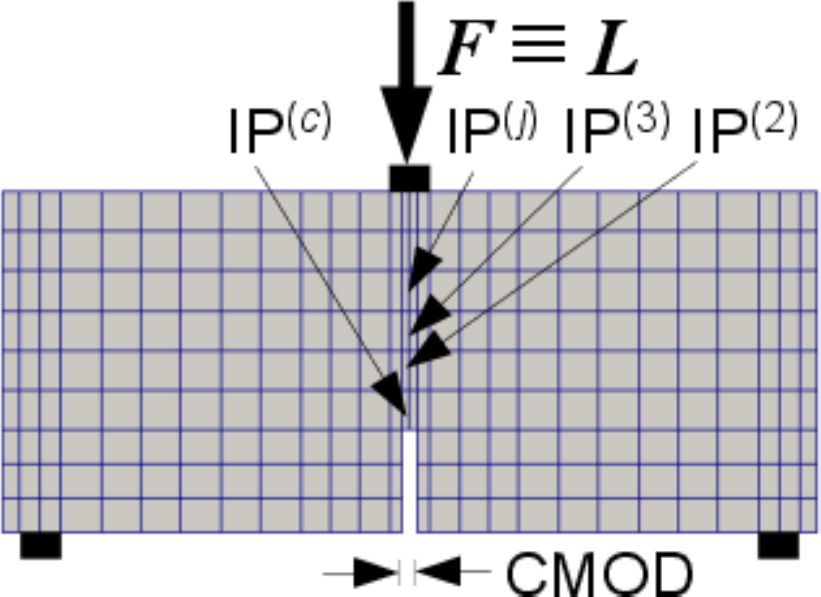} &
  \includegraphics[scale=0.35]{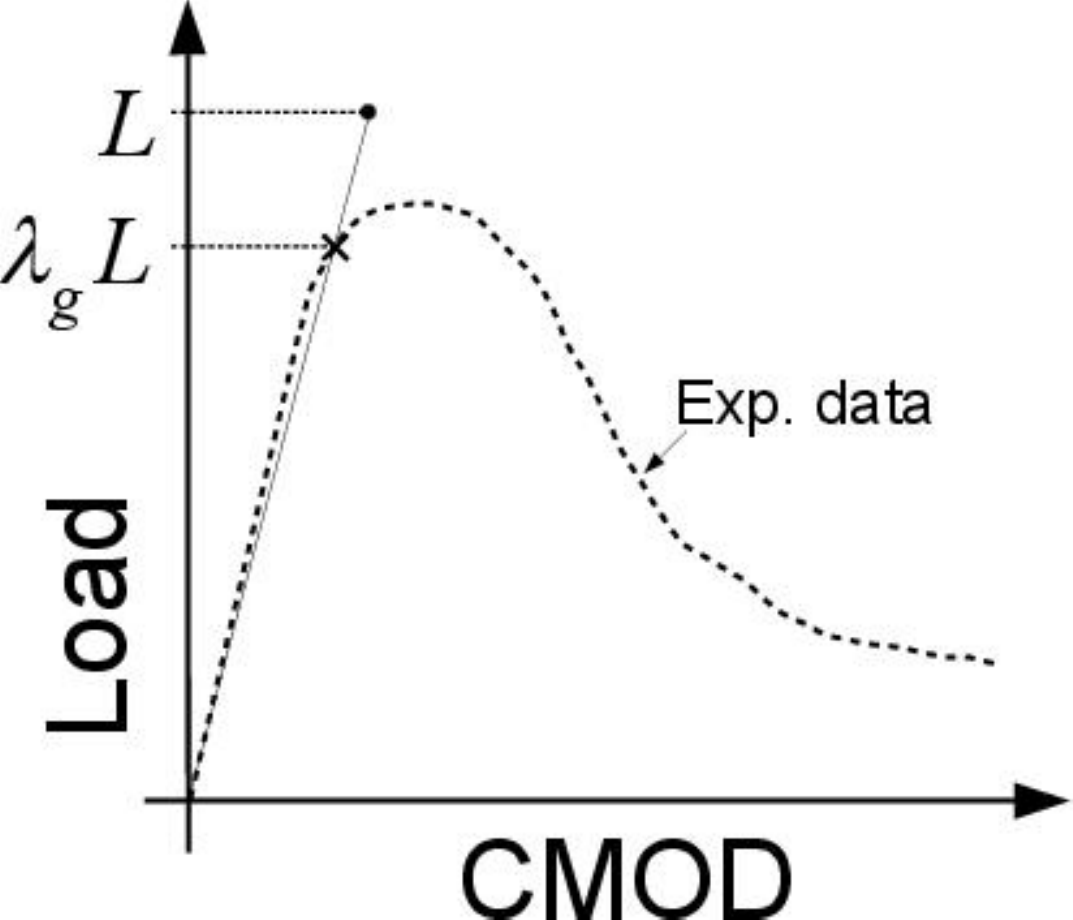} &
  \includegraphics[scale=0.35]{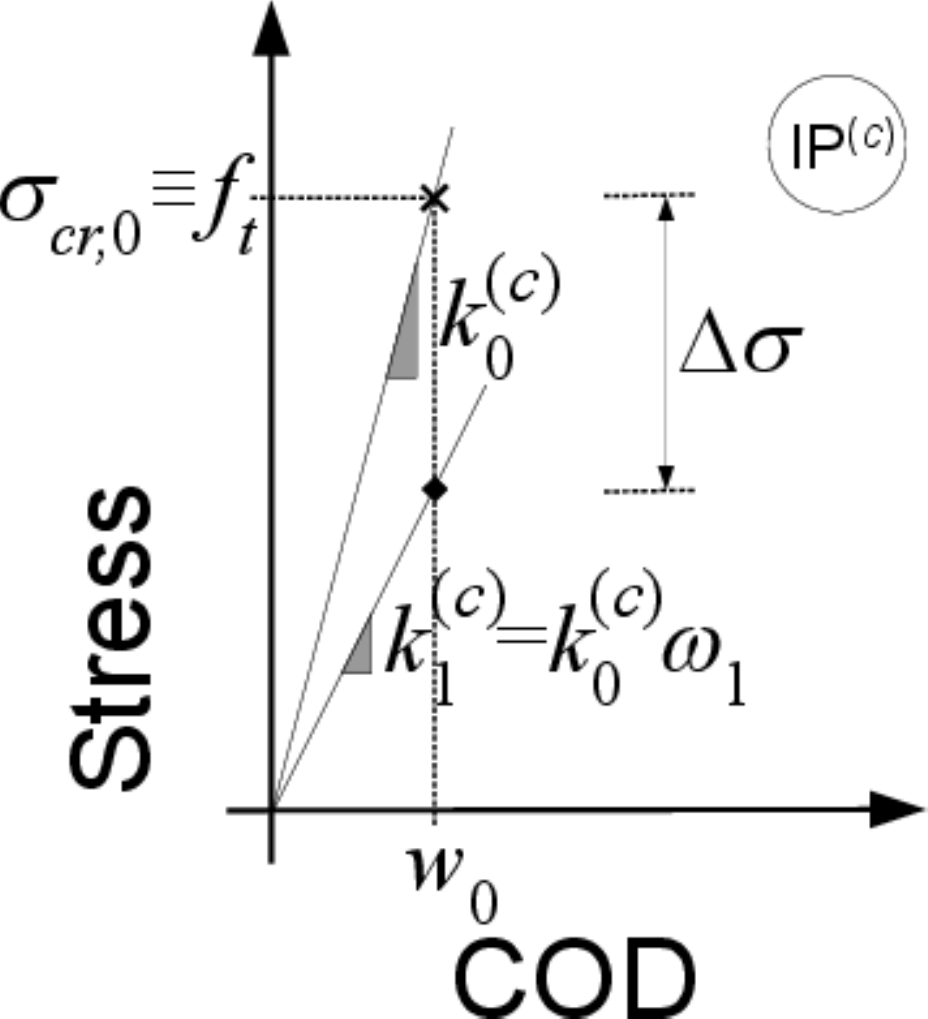} \\
  (a) & (b) & (c) \\
\end{tabular}
\caption{Inverse analysis calculus in Step~(1): a)~model and labels of integration points; b)~evaluation of load factor; c)~identification of TS relation and stiffness reduction in the lead integration point (IP$^{(c)}$)}
\label{fig:IA_1}
\end{center}
\end{figure}

Step~(1): In the first step, a reference load $L$ is applied to the FE model~(\figref{fig:IA_1}(a)). This load should be larger than the maximum load observed in the experiment. As the model is linear elastic, the load-displacement curve is a straight line. The load and displacement are scaled by a load factor $\lambda_g$, which is determined from the intersection of the calculated and experimental load-displacement curves (\figref{fig:IA_1}(b)). With the scaled load, the normal stress and the \newtext{COD} \oldtext{displacement jump} are calculated in integration points of all interface elements along the potential crack path. The integration point with the highest stress is identified as the point, which should crack in the current step. See \figref{fig:IA_1}(a), where it is denoted as IP$^{(c)}$. The stress and the \newtext{COD} \oldtext{displacement jump} in this integration point mark the first point of the identified TS relation $[w_0, \sigma_{cr,0}]$ (\figref{fig:IA_1}(c)). Note that $\sigma_{cr,0}$ corresponds to the tensile strength $f_t$ and $w_0$ is the small \newtext{COD} \oldtext{displacement jump} due to the finite initial stiffness of interface elements. In the forthcoming text, we will call IP$^{(c)}$, i.e. the integration point, which is used to determine the foremost point of the TS relation, the ``lead IP''. In preparation for the next step, the normal stiffness at the lead integration point IP$^{(c)}$ is reduced by factor $\omega_1$ (\figref{fig:IA_1}(c)):
\begin{equation} \label{eq:k2}
k_1^{(c)} = k_0^{(c)}\cdot\omega_1,
\end{equation}
where the index $\bullet^{(c)}$ indicates association with the lead integration point and
\begin{equation} \label{eq:omega1}
\omega_1 = \frac{\sigma_{cr,0} - \Delta\sigma}{\sigma_{cr,0}}.
\end{equation}
The value of parameter $\Delta\sigma$ has to be specified by \newtext{the} \oldtext{a} user and corresponds to the maximum allowable stress decrement between the individual points of the identified TS relationship and thus controls the stiffness reduction between individual calculation steps. At the same time, the strength $\sigma_{cr,0}\equiv f_t$ obtained in the first step is assigned to all integration points of interface elements along the potential crack path as their current strength.

%

\begin{figure}[ht]
\begin{center}
\begin{tabular}{ccc}
  \includegraphics[scale=0.32]{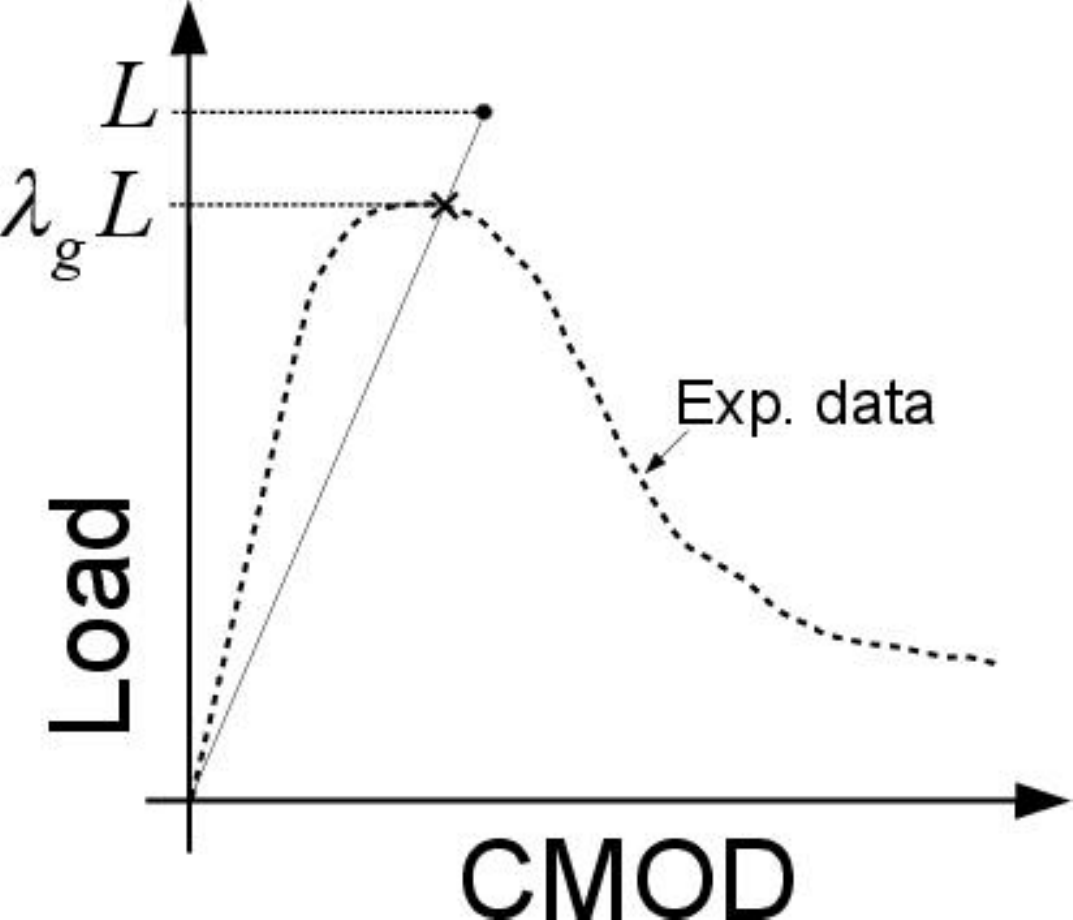} &
  \includegraphics[scale=0.32]{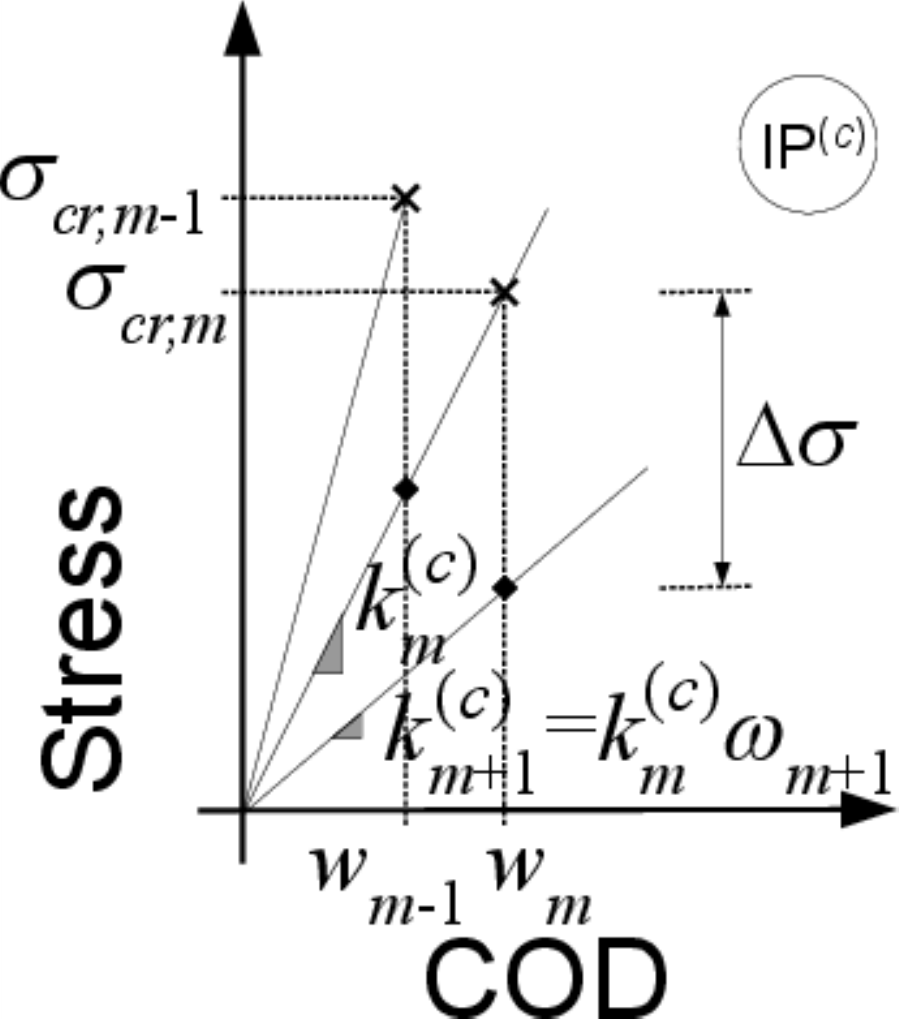} &
  \includegraphics[scale=0.32]{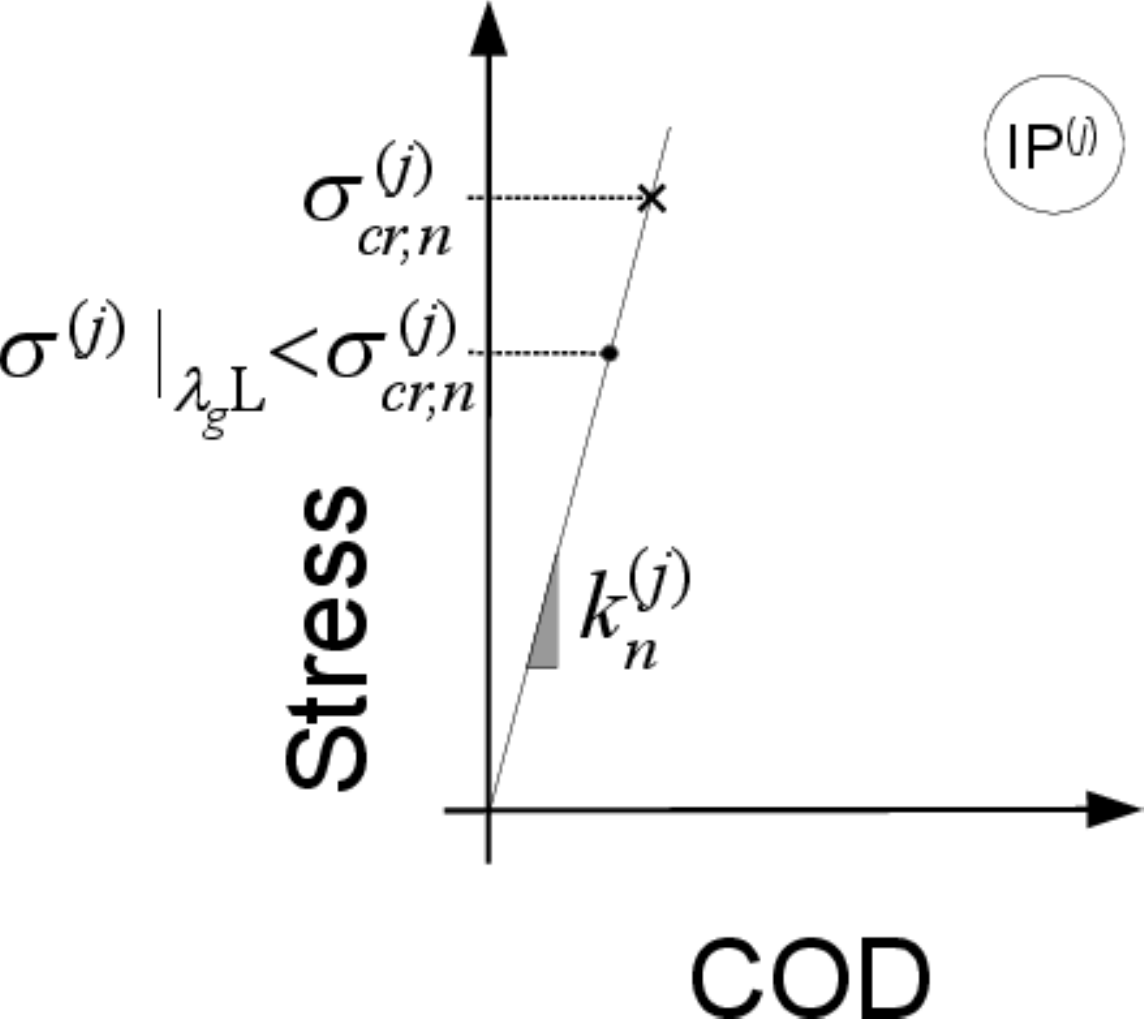} \\
  (a) & (b) & (c) \\
  \includegraphics[scale=0.32]{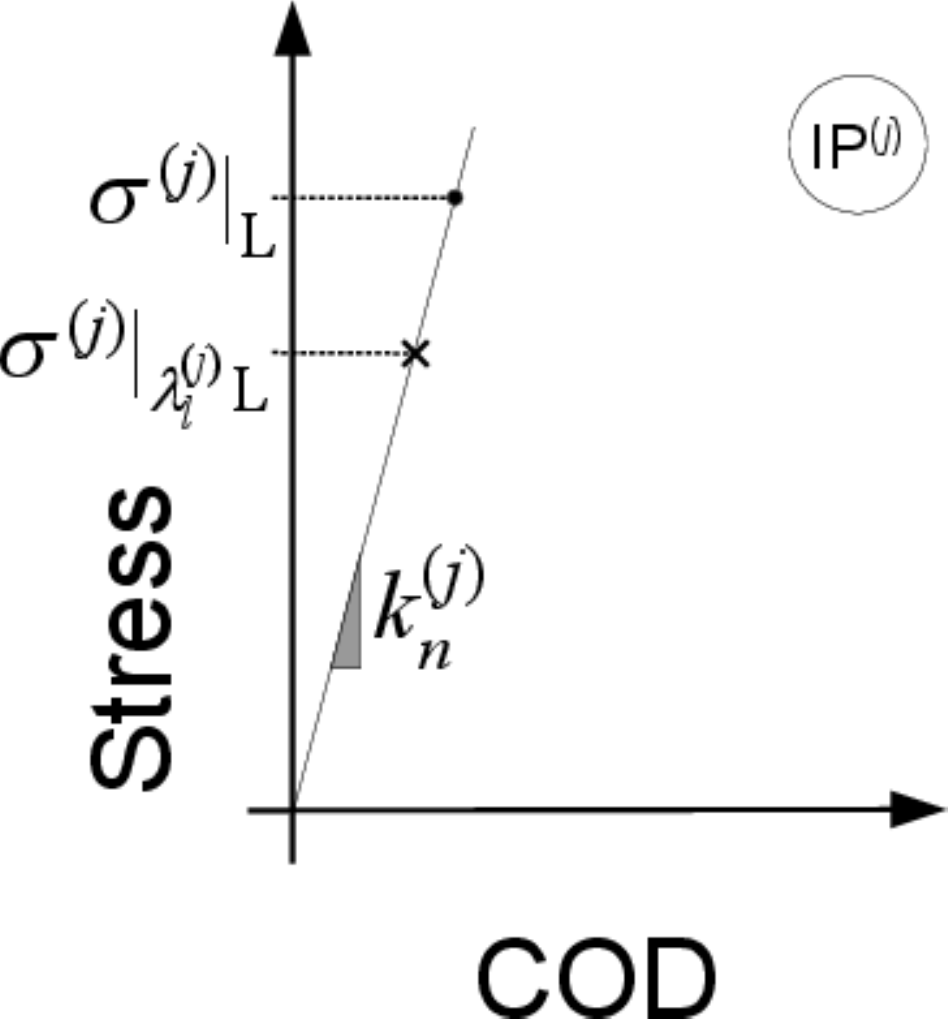} &
  \includegraphics[scale=0.32]{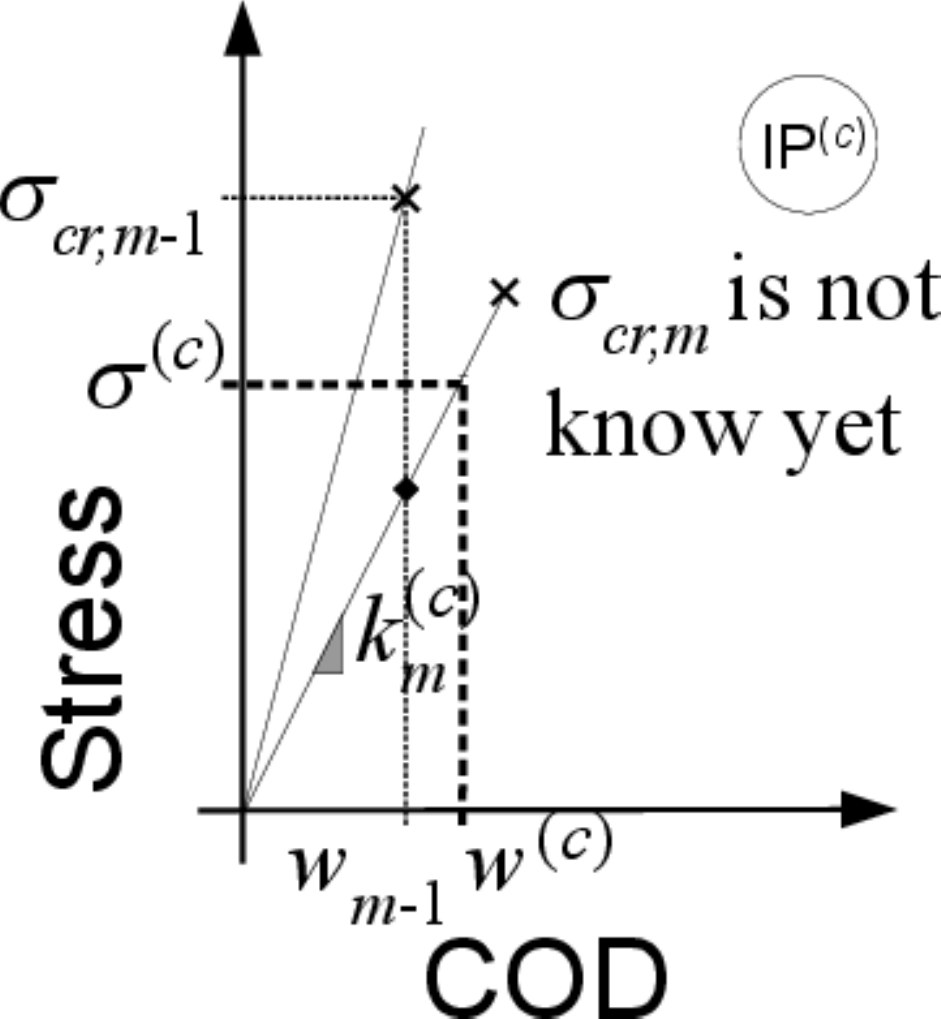} &
  \includegraphics[scale=0.32]{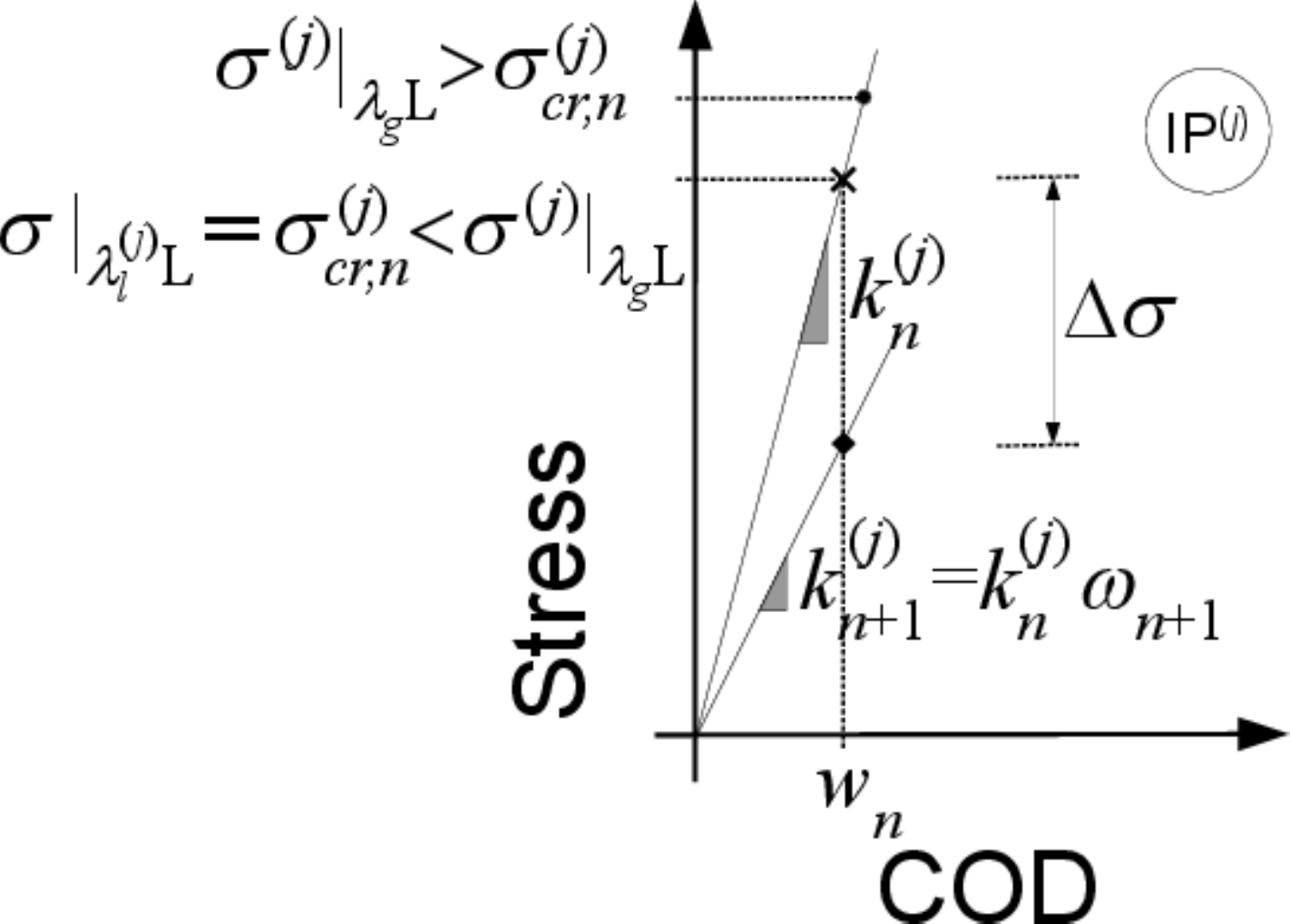} \\
  (d) & (e) & (f) \\
\end{tabular}
\caption{Inverse analysis calculus in \newtext{general Step~(i)} \oldtext{Step~(2)}: a)~evaluation of global load factor; b)~Case~(A) -~identification of TS relation and stiffness reduction in the lead integration point; c)~Case~(A) -~state in other integration points; d)~evaluation of local load factor; e)~Case~(B) - state in the lead integration point; f)~Case~(B) -~reduction of stiffness in other critical integration point}
\label{fig:IA_2}
\end{center}
\end{figure}

General Step~(i): Let us assume, that from the previous step(s), $m$ points of the TS relation $\sigma_{cr}(w)$ have been determined (note that indexing starts from 0). Recall that this relation specifies the maximum stress that any integration point on the existing and potential crack path can sustain depending on the magnitude of its \newtext{COD} \oldtext{displacement jump} $w$. At the same time, each integration point follows a linear relation between the normal cohesive stress and \newtext{COD} \oldtext{displacement jump} with current, possibly reduced, secant stiffness $k^{(j)}_n$
\begin{equation}
  \sigma^{(j)}(w) = k^{(j)}_n w^{(j)}
\end{equation}
The index $\bullet^{(j)}$ indicates association with the integration point $j$ and $n\leq m$ stands for the current $n$-th reduction step in a given integration point. Note that $n$ may be different in each integration point. The FE model is loaded by the reference load~$L$. However, besides determining the ``global'' load factor $\lambda_g$ from the intersection of the model and experimental load-displacement curves (\figref{fig:IA_2}(a)), ``local'' load factors $\lambda_{l}^{(j)}$ are calculated for each integration point on the existing and potential crack path, except the lead integration point IP$^{(c)}$. The local load factor is a value by which the normal stress ($\sigma^{(j)}$) in the integration point, due to the reference load $L$, would have to be scaled to become equal to the current critical strength ($\sigma_{cr,n}^{(j)}$) of the integration point and is evaluated as (\figref{fig:IA_2}(d))
\begin{equation} \label{eq:lambdal_cond}
\sigma^{(j)}(w)\rvert_{\lambda_{l}^{(j)}\,L} \equiv \lambda_{l}^{(j)}\,\sigma^{(j)}(w)\rvert_{L} = \sigma_{cr,n}^{(j)}
\end{equation}
Notation $\bullet|_{L}$ means that a variable is evaluated for the load $L$. The actual load factor for the present step $\lambda$ is found as the minimum:
\begin{equation} \label{eq:lambda}
\lambda=\min_{j}(\lambda_g,\lambda_{l}^{(j)}).
\end{equation}
The load is scaled with the factor $\lambda$ and then $\sigma$ and $w$ are updated in all interface elements using their current stiffness. Interpretation of the present step then depends on which of the load factors is determinant:
\begin{itemize}
  \item Case~(A), when $\lambda = \lambda_g$, see Figs.~\ref{fig:IA_2}(b,c): In this case, the calculation reached another point on the experimental load-displacement curve. The integration point with the lowest current stiffness is selected as the lead integration point IP$^{(c)}$. For an experiment with monotonously propagating and opening crack, this integration point is the same as in the Step~(1), but, in general, if multiple integration points have this lowest stiffness, the integration point with the highest stress is identified as the point, which should crack in the current step. The scaled \newtext{COD} \oldtext{displacement jump} and stress at the lead integration point identify a new point of the TS relationship $[w_{m},\sigma_{cr,m}]$. The stiffness of this integration point is reduced using the parameter $\Delta\sigma$, analogously with Eqs.\,\eqref{eq:k2} and~\eqref{eq:omega1}, as
      \begin{equation} \label{eq:km}
      k_{m+1}^{(c)} = k_m^{(c)}\cdot\omega_{m+1},~~\omega_{m+1} = \frac{\sigma_{cr,m} - \Delta\sigma}{\sigma_{cr,m}}.
      \end{equation}
  \item Case~(B), when $\lambda = \lambda_{l}^{(j)}$, see Figs.~\ref{fig:IA_2}(e,f): This means that the integration point number $j$ should crack, but the load-displacement state at which it cracks is not yet on the measured load-displacement curve. This is an intermediate state, at which no new point of the TS relation is determined. For the further calculation, the secant stiffness of this integration point is reduced using the parameter $\Delta\sigma$, analogously with Eqs.\,\eqref{eq:k2},\eqref{eq:omega1} and \eqref{eq:km},
      and the new corresponding current strength is assigned according to the already determined TS diagram.
\end{itemize}

\begin{figure}
\begin{center}
\begin{tabular}{c}
\scalebox{0.8}{
\tikzstyle{box} = [rectangle, rounded corners, minimum width=3cm, minimum height=0.5cm,text centered, draw=black, fill=white!30]
\tikzstyle{io} = [trapezium, trapezium left angle=50, trapezium right angle=130, minimum width=1cm, minimum height=0.5cm, text centered, draw=black, fill=gray!30]
\tikzstyle{process} = [rectangle, minimum width=3.5cm, minimum height=0.5cm, text centered, draw=black, fill=gray!20]
\tikzstyle{process1} = [rectangle, minimum width=3.5cm, minimum height=0.5cm, text centered, draw=black, fill=gray!40]
\tikzstyle{process2} = [rectangle, minimum width=3.5cm, minimum height=0.5cm, text centered, draw=black, fill=gray!60]
\tikzstyle{decision} = [diamond, aspect=2, minimum width=3cm, minimum height=0.5cm, text centered, draw=black, fill=white!30]
\tikzstyle{point} = [coordinate]

\tikzstyle{arrow} = [thick,->,>=stealth]
\tikzstyle{line} = [thick,-,>=stealth]

\tiny
\begin{tikzpicture} [ node distance = 1.2cm, thick, nodes = {align = center}, >=latex]

  \matrix [column sep=5mm,row sep=3mm] {
    &\node (Start) [box] {Start of analysis}; \\
    &\node (Sload) [process] {Reference load $L$ is applied}; \\
    &\node (Sss) [process] {Calculation of stress-strain state \\ for load $L$}; \\
    &\node (Slam1) [process] {Calculation of load factors $\lambda_g$ and $\lambda_{l}^{(j)}$}; \\
    &\node (Slam2) [process] {$\lambda=\min\limits_{j}(\lambda_g,\lambda_{l}^{(j)})$}; \\

    \node (Ati) [process1] {Continuation of inverse analysis}; &
    \node (Ldecision) [decision] {Is $\lambda$ found?}; \\

    \node (Adecision) [decision] {$\lambda_g < \min\limits_{j}(\lambda_{l}^{(j)})$}; &&
    \node (Iexit) [box] {EXIT}; \\

    \node (Alcur11) [process1] {Cracking of IP$^{(j)}$}; &
    \node (Alcur21) [process1] {Cracking of IP$^{(c)}$}; & \\

    \node (Alcur12) [process1] {Stiffness reduction \\ and assignment of new $\sigma_{cr}^{(j)}$ for IP$^{(j)}$  }; &
    \node (Alcur22) [process1] {Stiffness reduction in IP$^{(c)}$, \\ new point of TS relationship }; \\

    \node (Aps)  [process1] {Update of global stiffness matrix, \\ data saving }; \\
  }; 
  \draw[arrow]  (Start)      -- (Sload);
  \draw[arrow]  (Sload)      -- (Sss);
  \draw[arrow]  (Sss)        -- (Slam1);
  \draw[arrow]  (Slam1)      -- (Slam2);
  \draw[arrow]  (Slam2)      -- (Ldecision);

  \draw[arrow]  (Ldecision)  -- node[anchor=south]{YES} (Ati);
  \draw[arrow]  (Ldecision)  -| (Iexit);
  \node (dummy2) [right of=Ldecision, xshift=0.7cm, yshift=0.15cm] {NO};
  \node (dummy3) [right of=Ldecision, xshift=1.3cm, yshift=-0.5cm,align=left] {Missing input data \\ or end of loading \\ curve was reached};

  \draw[arrow]   (Ati)       -- (Adecision);
  \draw[arrow]  (Adecision)  -- node[anchor=east]{NO}  (Alcur11);
  \node (dummy5) [below of = Adecision, xshift=0.5cm, yshift=0.34cm] {Case (B)};
  \draw[arrow]  (Adecision)  -| (Alcur21);
  \node (dummy1) [right of=Adecision, xshift=0.9cm, yshift=0.15cm] {YES};
  \node (dummy4) [right of=Adecision, xshift=0.9cm, yshift=-0.15cm] {Case (A)};

  \draw[arrow]   (Alcur11)   -- (Alcur12);
  \draw[arrow]   (Alcur21)   -- (Alcur22);

  \draw[arrow]   (Alcur12)   -- (Aps);
  \draw[arrow]   (Alcur22)   |- (Aps);

  \draw[arrow]  (Aps)        --  ++(-2.0,0) |- (Sss);

\end{tikzpicture}
}
\end{tabular}
\caption{Flowchart of the inverse analysis based on SLA}
\label{fig:flow}
\end{center}
\end{figure}
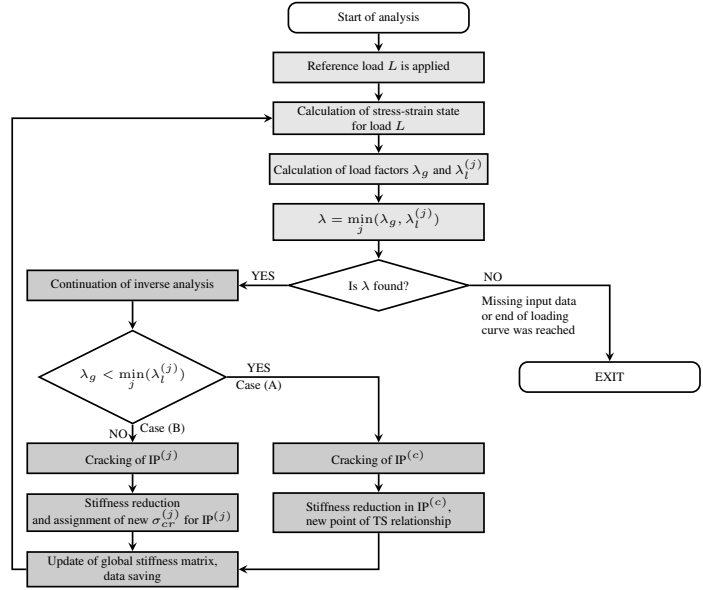

The above-described step is repeated until the experimental load-displacement curve is completely traced, see also the procedure flowchart in~\figref{fig:flow}. Note that in the subsequent examples, all points of the calculated load-displacement curve corresponding to Cases~(A) and (B) are plotted. Nevertheless, only points corresponding to Cases~(A) represent states that are used to retrieve the TS relation. \newtext{It has to be noted that the obtained TS diagram is often uneven and the same approach presented in~\cite{Nanakorn:1996:Back} can be employed to smooth out the curve for a later utilisation in FE simulations. More specifically, the curve can be smoothed in such a way that the area under the smoothed curve is equal to the area under the original TS diagram. This method can be seen as the reverse approach to the derivation of the saw-tooth law from a smooth TS diagram, see Section~\ref{sec:sla} and \figref{fig:SLA}(b).}

A few limitations of the proposed inverse analysis should be noted:
\begin{itemize}
\item Initial stiffness of the numerical model must be equal or less than the measured stiffness.
\item If the experimental data do not extent up to the state when a traction-free crack starts to propagate, the complete TS relation cannot be retrieved.
\item The accuracy of the inverse analysis result depends on the accuracy of measured data.
\end{itemize}

\subsection{Multi-pass enhancement}
\label{subsec:multipass}
It is obvious that in the above-described procedure, the lead integration point IP$^{(c)}$ is used to identify all points of the TS relationship. Fracture tests are usually configured in such a way that one crack propagates and opens in a monotonous way. Then the integration point which fractures first and subsequently maintains the largest COD is the lead integration point IP$^{(c)}$. For example, in a bending test on a notched beam, the IP$^{(c)}$ is located at the tip of the notch. Once the cohesive stress in the lead IP$^{(c)}$ reaches zero (or the COD reaches the critical value of $w_{c}$), the TS law is completely defined. This TS relationship is then used in all subsequently fracturing elements along the crack path. However, depending on the experimental configuration, ligament size and material brittleness, the recorded loading curve may include the specimen response even beyond the state when the first lead IP$^{(c)}$ becomes traction-free. In other words, the data may capture propagation of a traction-free crack. In order to systematically utilise this additional information, we propose the following multi-pass enhancement of the inverse analysis.

In the first pass, the method is employed, as described in Section~\ref{subsec:principle}, to determine the TS relationship using the first lead integration point. Once the TS relation is completely identified, the first pass is terminated. If the input load-displacement curve extends beyond this state, the second pass is executed. In the second pass, the previously identified TS law is assigned to the first lead integration point. The inverse analysis is run again, but the integration point with the second largest COD is used to determine a new TS relation, which is then used for all subsequently cracking elements in this pass. Further passes are run until the end of the available loading curve is reached. As a result, as many TS relations as there were the passes are obtained. The difference of these TS relations to some extent reflects the inherent heterogeneity of the ligament through which the crack propagates. For further use in predictive analyses, these TS relations may be averaged to obtain a single material characteristic. Alternatively, the scatter of the obtained TS relations could be used in the context of stochastic FE analysis.

As we will see in the forthcoming examples (Section~\ref{sec:numSim}), the proposed multi-pass enhancement reduces the oscillations of the obtained TS relations and improves the accuracy. It is also shown, that when the TS relation is used to reproduce the experiment (for verification) the complete loading curve is matched.

\section{Numerical simulations}
\label{sec:numSim}
\oldtext{In this section we perform a sensitivity study of the proposed method and demonstrate its application to different materials and experimental configurations.} \newtext{A sensitivity study is carried out to ascertain the performance of the proposed method, folowed by its application to different materials and experimental configurations.} The method has been implemented for plane stress problems in the open source object oriented finite element program \code{OOFEM}~\cite{PatBit01}. In the forthcoming examples, isoparametric four-node quadrilateral plane-stress finite elements and cohesive interface elements with a linear approximation of \newtext{the} displacement field are used for the finite element calculations. The default element size is chosen equal to 10\,mm for the sensitivity study and 5\,mm for the other examples. If not defined differently, the parameter $\Delta\sigma$, controlling the inverse analysis, is set to be equal to 1\% of the determined tensile strength.

\subsection{Sensitivity study}
\label{subsec:sensitivityStudy}
\begin{figure}
\begin{center}
\begin{tabular}{cc}
  \includegraphics[width=0.4\textwidth]{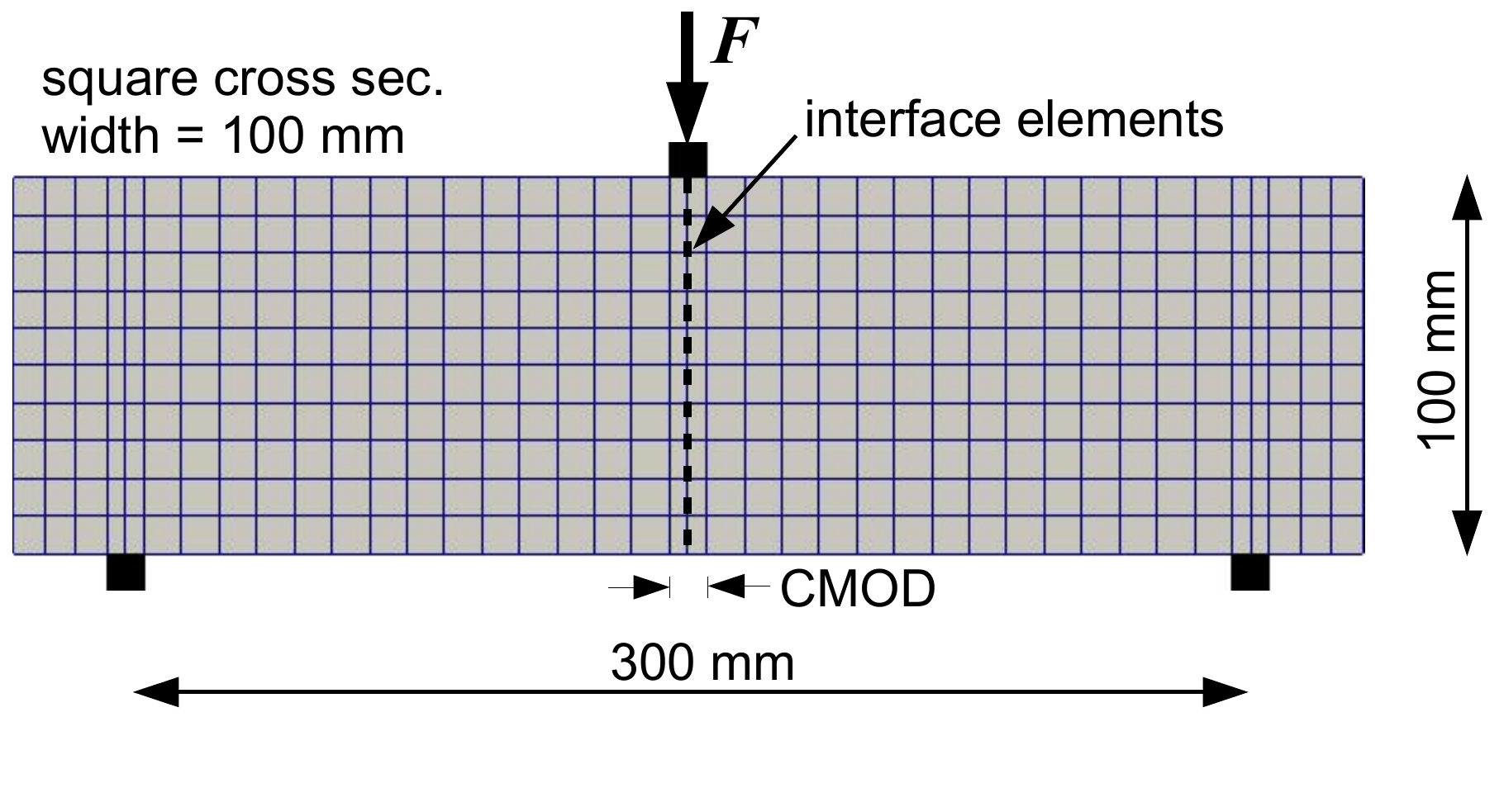} &
  \includegraphics[width=0.4\textwidth]{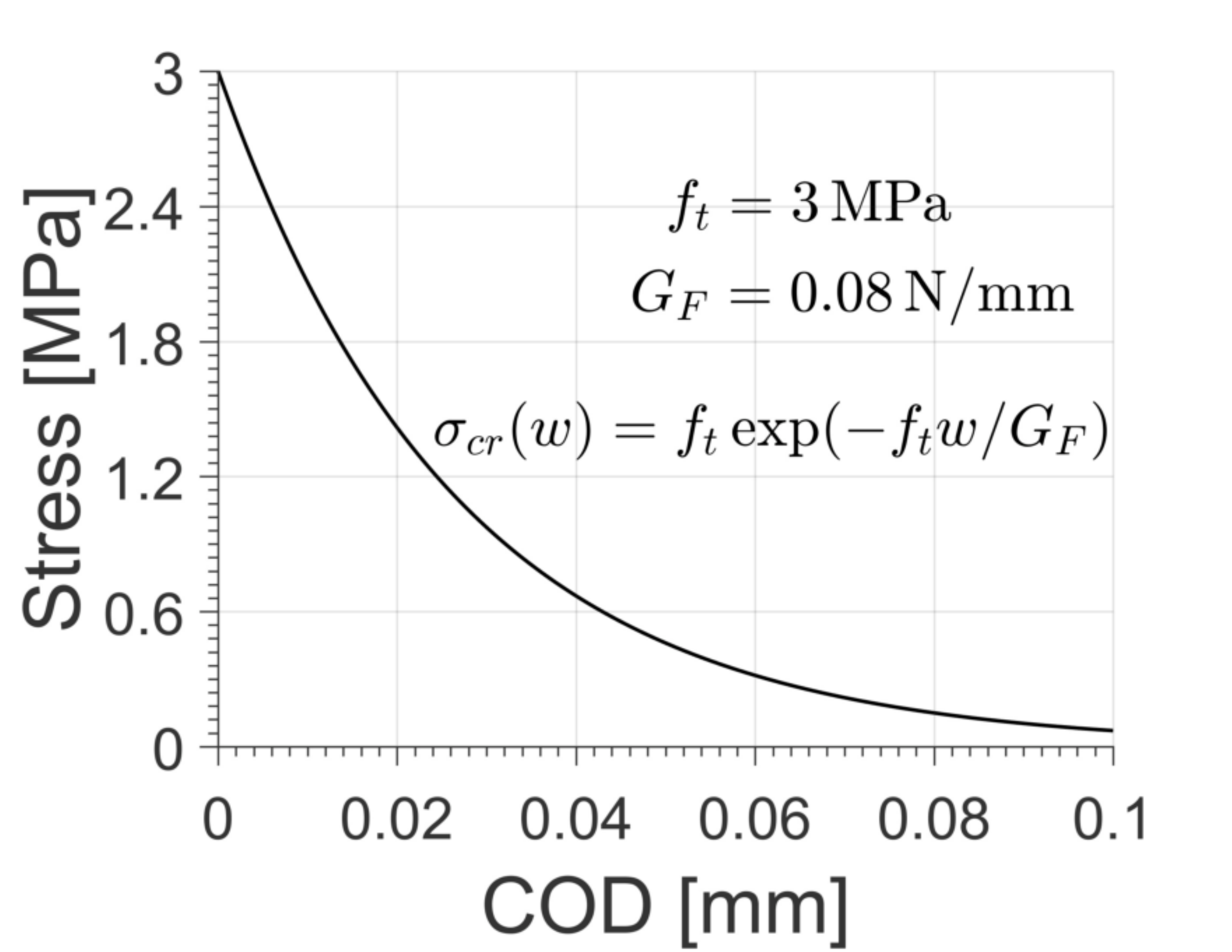} \\
  (a) & (b)
\end{tabular}
\caption{Three-point bending test used for the sensitivity study: a)~discretization; b)~exponential softening}
\label{fig:3PBtestSetup}
\end{center}
\end{figure}

\begin{figure}
\begin{center}
\begin{tabular}{cc}
  \includegraphics[width=0.4\textwidth]{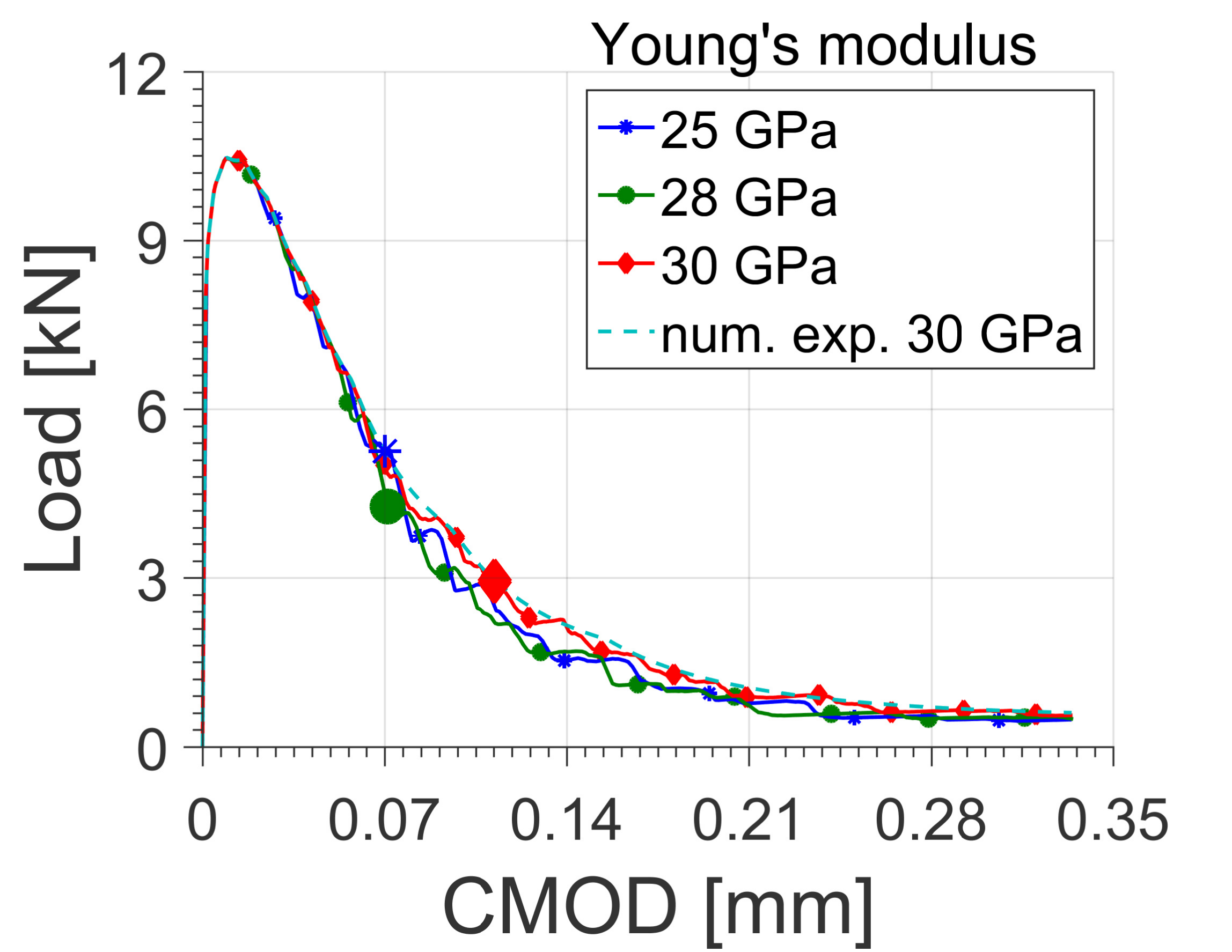} &
  \includegraphics[width=0.4\textwidth]{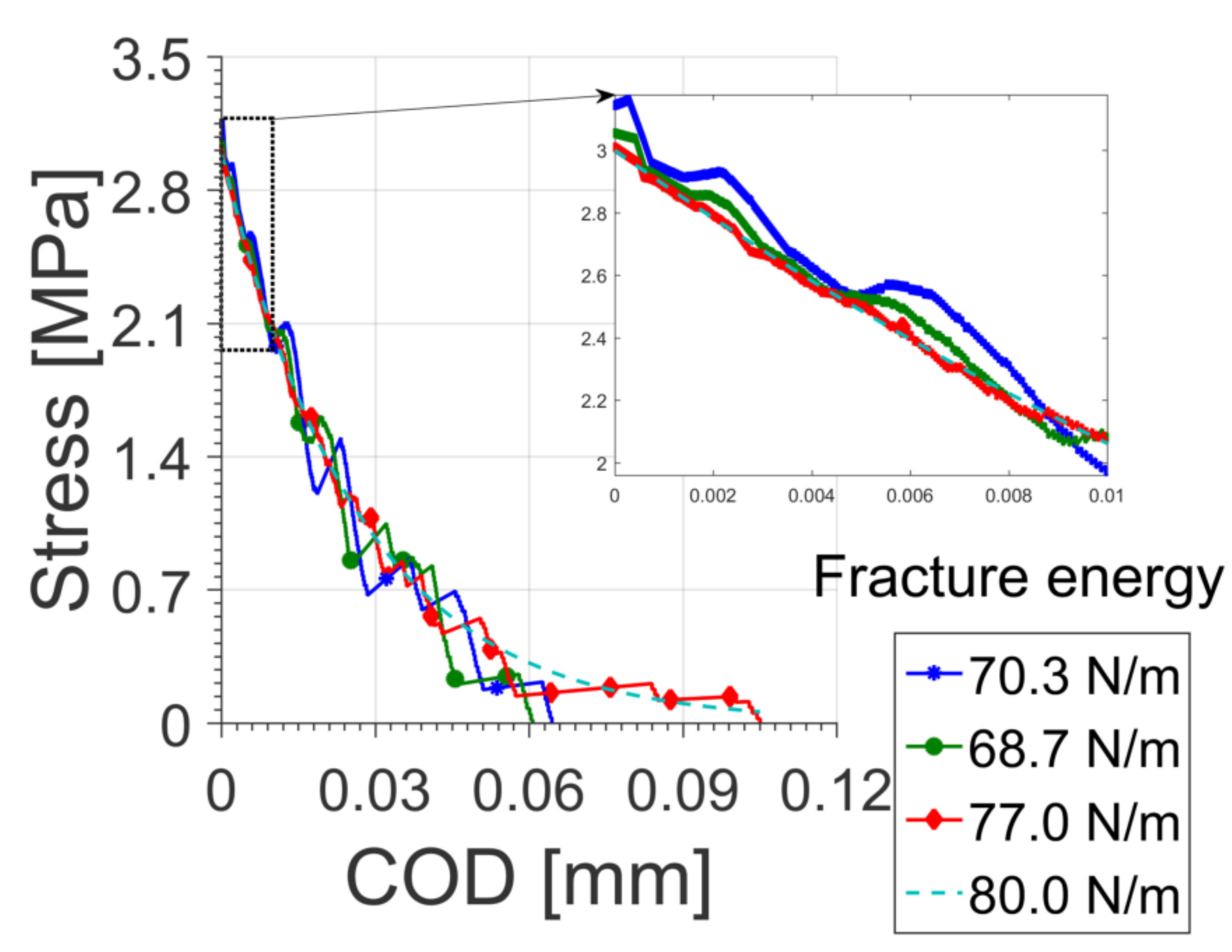} \\
  (a) & (b)
\end{tabular}
\caption{Three-point bending test - varying Young's modulus: a)~load-CMOD diagrams; b)~TS diagrams and corresponding fracture energies}
\label{fig:3PBtest_E}
\end{center}
\end{figure}

\oldtext{This section is devoted to the verification of the proposed method and the study of result dependence on the choice of  parameters specified by the user, i.e. Young's modulus, $\Delta\sigma$ and element size, see~\figsref{fig:3PBtest_E}{fig:3PBtest_elemSize}.} \newtext{To verify the proposed method and variation of results caused by the choice of user defined parameters, i.e. the Young modulus of bulk material, $\Delta\sigma$ and element size, a sensitivity study is presented hereafter.} The influence of initial normal and shear stiffnesses of interface elements ($k_0,g_0$) is not presented herein because they serve as the penalty parameters and do not generally influence the obtained results. As the input, the loading curve obtained from an artificial ``numerical'' experiment simulating a three-point bending test without \oldtext{the} notch (\figref{fig:3PBtestSetup}(a)) is used. The beam is modelled by means of a linear elastic isotropic material characterised by Young's modulus $E = 30$\,GPa and Poisson's ratio $\nu=0.2$. Fracture is represented by cohesive interface elements with an exponential TS relation, where $f_t=3$\,MPa and the total fracture energy $G_F=80$\,N/m (\figref{fig:3PBtestSetup}(b)).

For all study cases the load-CMOD diagrams as well as the tension TS diagrams with corresponding values of fracture energy are presented \newtext{in the following figures} (dashed line represents the artificial ``numerical'' experiment and the solid line indicates the inverse analysis). The presented results clearly show the sensitivity of the current procedure to the varying initial parameters set up by the user. The bigger markers depicted in the load-CMOD diagrams represent the end of the inverse analysis. Note that the further response is driven only by the already determined TS diagram.

As already mentioned in Section~\ref{subsec:principle}, the Young modulus of bulk material has to be prescribed by the user. \oldtext{For the numerical examples presented in Section~\ref{sec:numStudy}, the value is always determined by matching the numerical response to the initial elastic part of input experimental data. Note that the elastic numerical response has to overlap or be below the experimental data, i.e. the assigned Young modulus is less or equal to the true one, otherwise the criteria for obtaining TS diagram would not be fulfilled, see Section~\ref{subsec:principle}.} \figref{fig:3PBtest_E} shows the results for three different values of Young's modulus, and corresponding TS diagrams. As can be seen, the method provides good match to known TS diagram even for the modulus value of 25\,GPa which is cca 17\% lower then the real one.

%

%

\begin{figure}
\begin{center}
\begin{tabular}{cc}
  \includegraphics[width=0.4\textwidth]{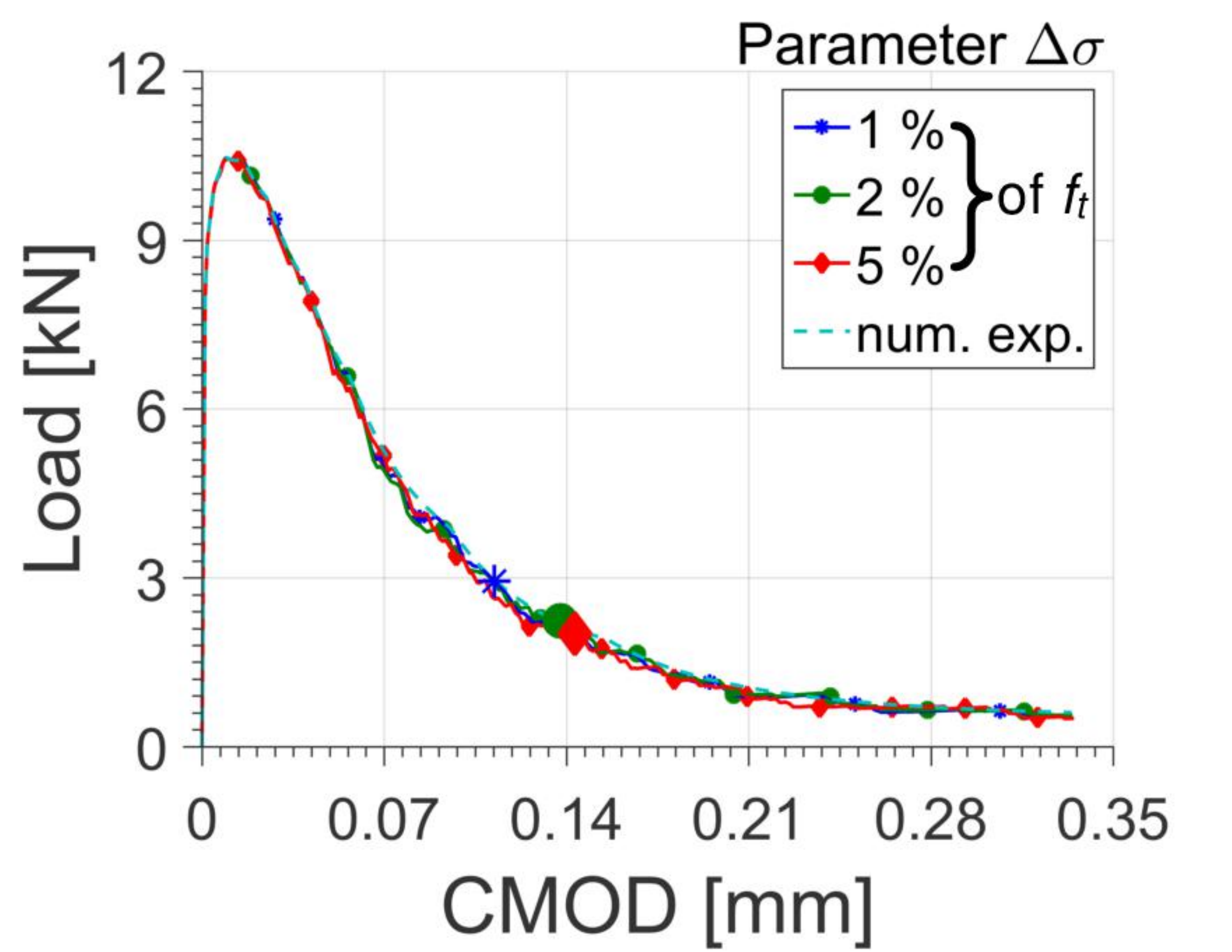} &
  \includegraphics[width=0.4\textwidth]{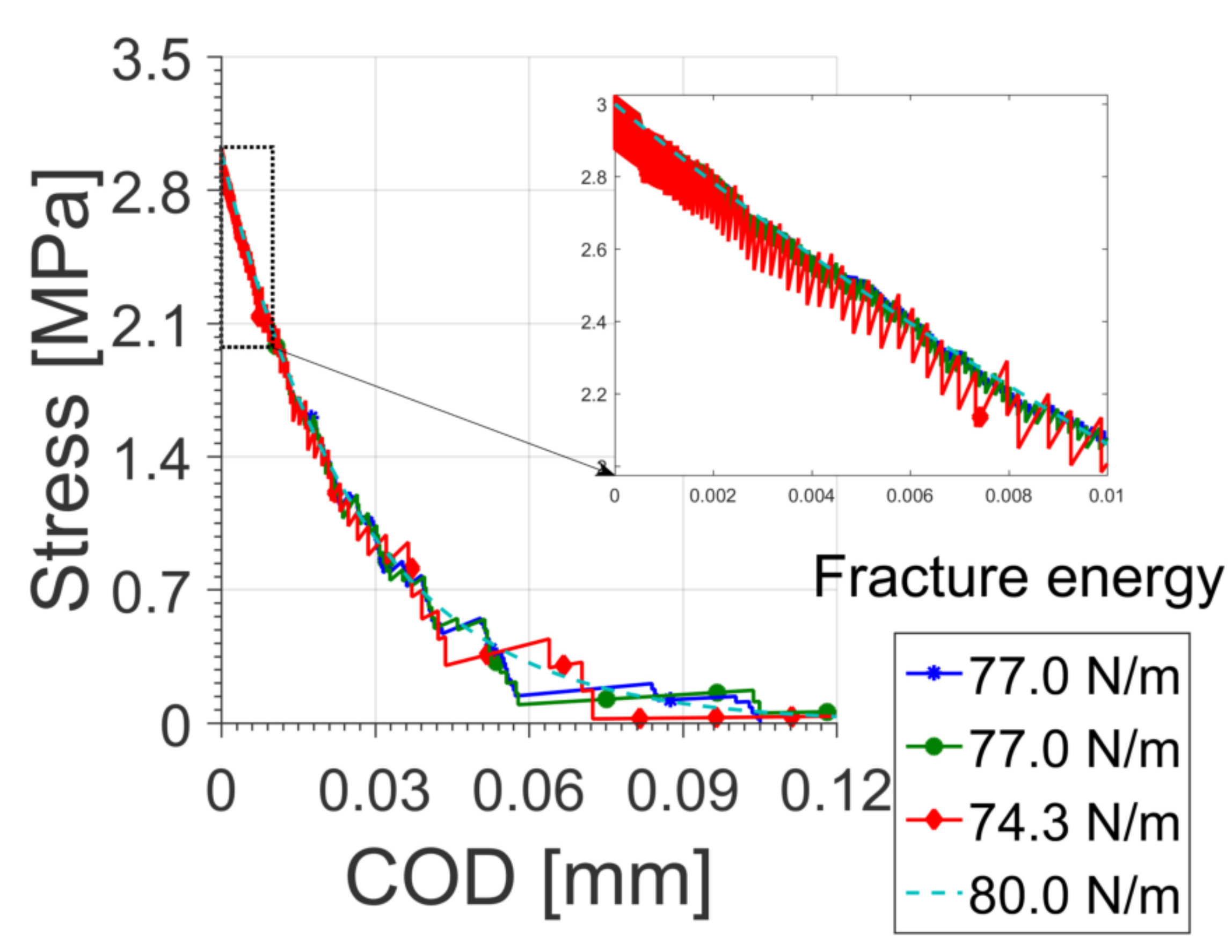} \\
  (a) & (b)
\end{tabular}
\caption{Three-point bending test - varying $\Delta\sigma$: a)~load-CMOD diagrams; b)~TS diagrams and corresponding fracture energies}
\label{fig:3PBtest_jumpPeak}
\end{center}
\end{figure}

The second example shows the variation of inverse analysis results for $\Delta\sigma$, which stands for the maximum allowable stress jump (prescribed as the percentage of determined tensile strength). The assumed values are $1,2,5$\% of $f_t$ and the results are shown in~\figref{fig:3PBtest_jumpPeak}. As can be seen in~\figref{fig:3PBtest_jumpPeak}(b), the main difference is in the smoothness of obtained TS diagrams, especially in the late post-peak, but overall good match is obtained.

\begin{figure}
\begin{center}
\begin{tabular}{cc}
  \includegraphics[width=0.4\textwidth]{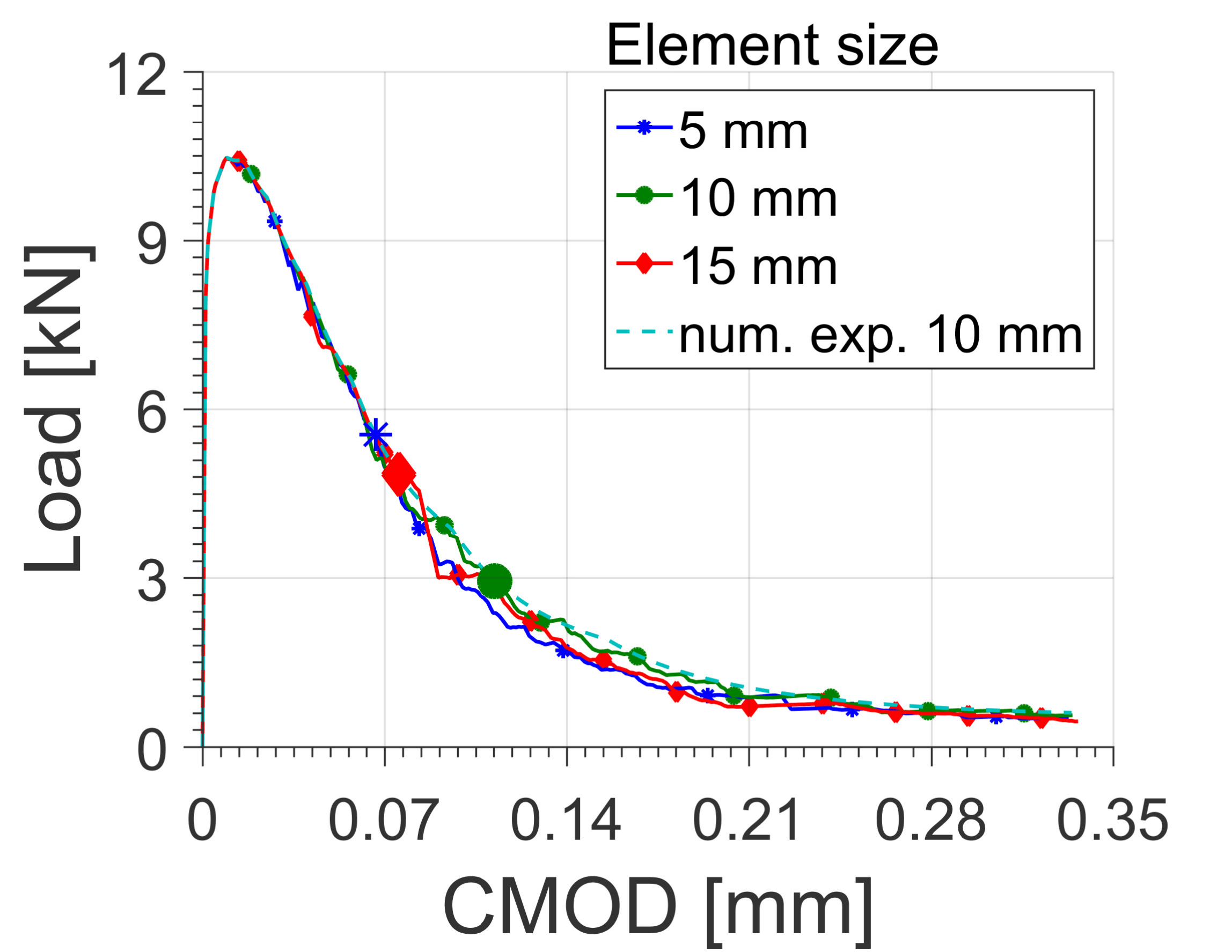} &
  \includegraphics[width=0.4\textwidth]{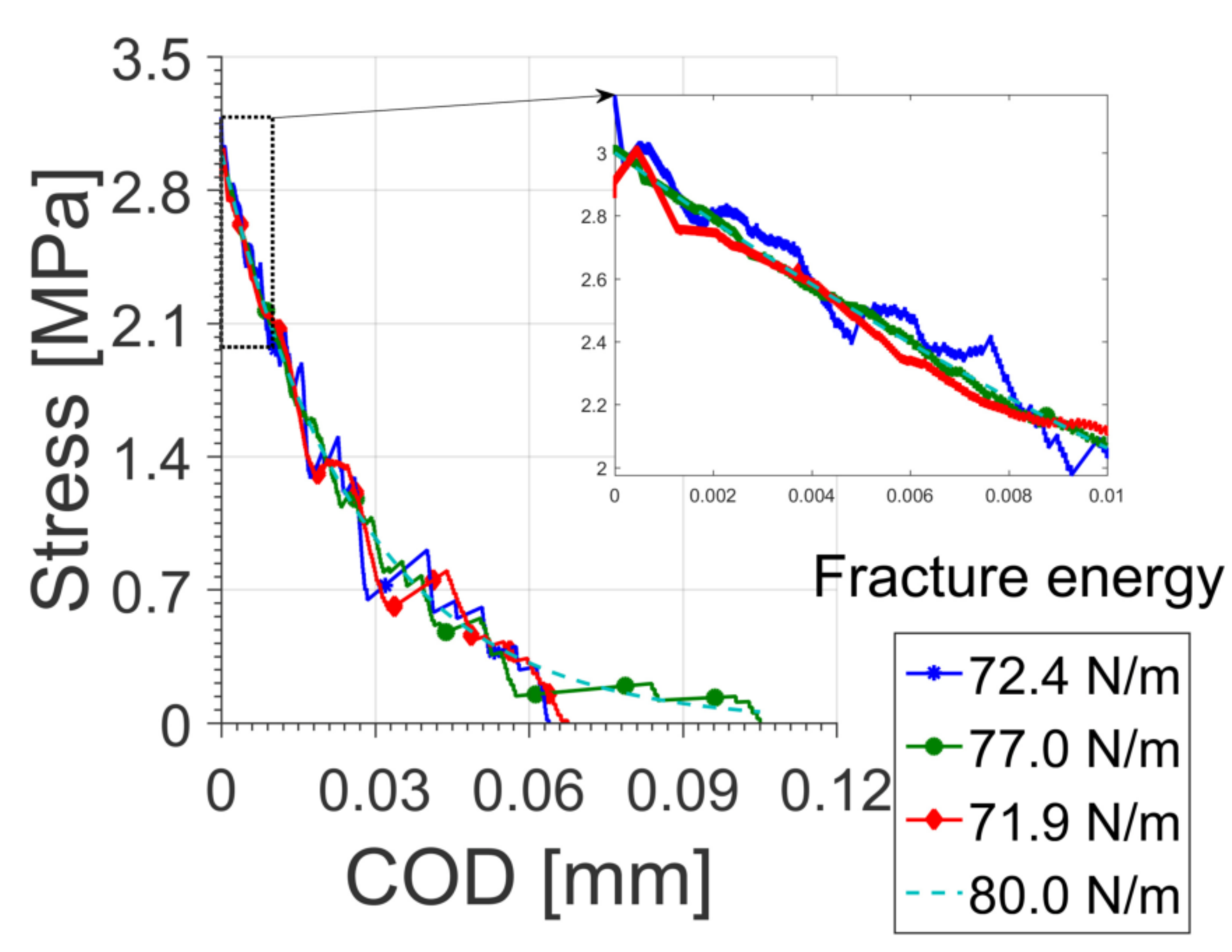} \\
  (a) & (b)
\end{tabular}
\caption{Three-point bending test - varying element size: a)~load-CMOD diagrams; b)~TS diagrams and corresponding fracture energies}
\label{fig:3PBtest_elemSize}
\end{center}
\end{figure}

\figref{fig:3PBtest_elemSize} is used to demonstrate the influence of utilised element size, i.e. the position of integration point driving the determination of TS diagram. Three different sizes $5,10,15$\,mm are assumed (the original size for artificial ``numerical'' experiment is equal to 10\,mm). \figref{fig:3PBtest_elemSize}(b) shows that all sizes provide results, which correspond well to the prescribed TS diagram and slightly deviate in the late post-peak.

As can be seen from the presented results (\figsref{fig:3PBtest_E}{fig:3PBtest_elemSize}) the procedure is capable to determine the TS diagram with sufficient accuracy even if the user-defined parameters are chosen in a broader range or with some error compared to the real properties.

\begin{figure}
\begin{center}
\begin{tabular}{cc}
  \includegraphics[width=0.4\textwidth]{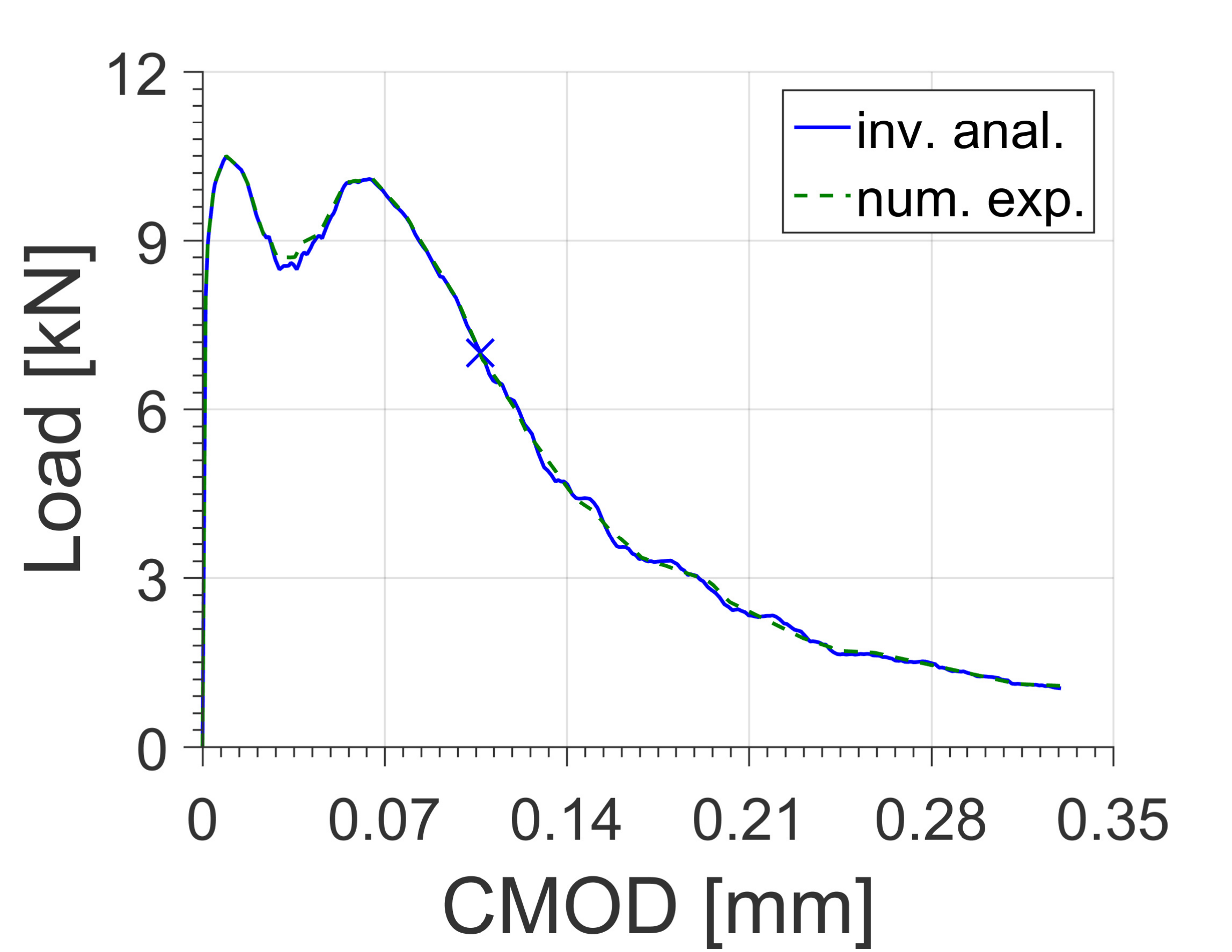} &
  \includegraphics[width=0.4\textwidth]{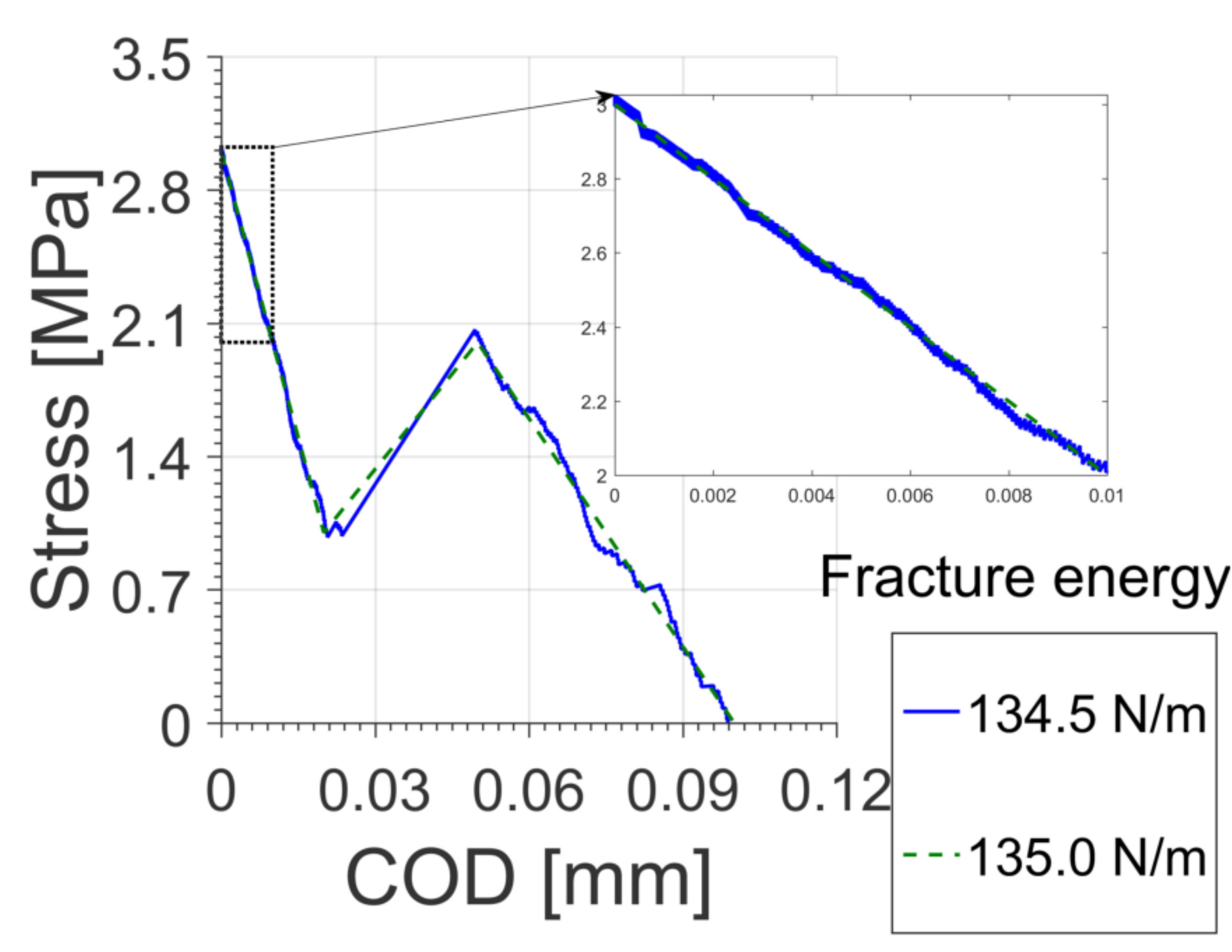} \\
  (a) & (b)
\end{tabular}
\caption{Three-point bending test: a)~load-CMOD diagrams; b)~TS diagrams and corresponding fracture energies}
\label{fig:3PBtest_hard}
\end{center}
\end{figure}

The last example in this section (\figref{fig:3PBtest_hard}) demonstrates the ability of the proposed inverse analysis to
treat load-displacement (CMOD) data with hardening interval in the post-peak phase. In this case, the TS relation was not monotonously softening, but included a hardening portion after the initial softening (\figref{fig:3PBtest_hard}(b)). As seen in~\figref{fig:3PBtest_hard}(b), the TS relation is very well reproduced by the inverse calculation. It should be noted, though, that in the present method, the linear stiffness of the model must monotonously decrease as the load-displacement curve is traced. Thus, it cannot be applied to those cases, when the slope of the hardening portion of the load-displacement data would be larger than the secant stiffness.

\subsection{Inverse analysis of real experimental data}
\label{sec:numStudy}

To demonstrate the versatility of the proposed inverse analysis, real experimental data for various materials are utilised to obtain the traction-separation diagrams. Similarly to Section~\ref{subsec:sensitivityStudy}, the load-CMOD and load-displacement diagrams as well as the tension TS diagrams with corresponding fracture energies are presented. In the load-displacement (CMOD) graphs, the dashed lines depict the experimentally obtained data which served as the input for the inverse analysis. The solid lines represent the reproduction of these data achieved during the inverse analysis. Note that the latter curves include not only the Cases~(A), when the model actually matched the experiment, but also the intermediate Cases~(B), when other points than the lead integration point cracked (see Section~\ref{subsec:principle}). The bigger markers indicate the points at which the inverse analysis was terminated.

\begin{figure}
\begin{center}
\begin{tabular}{cc}
  \includegraphics[width=0.4\textwidth]{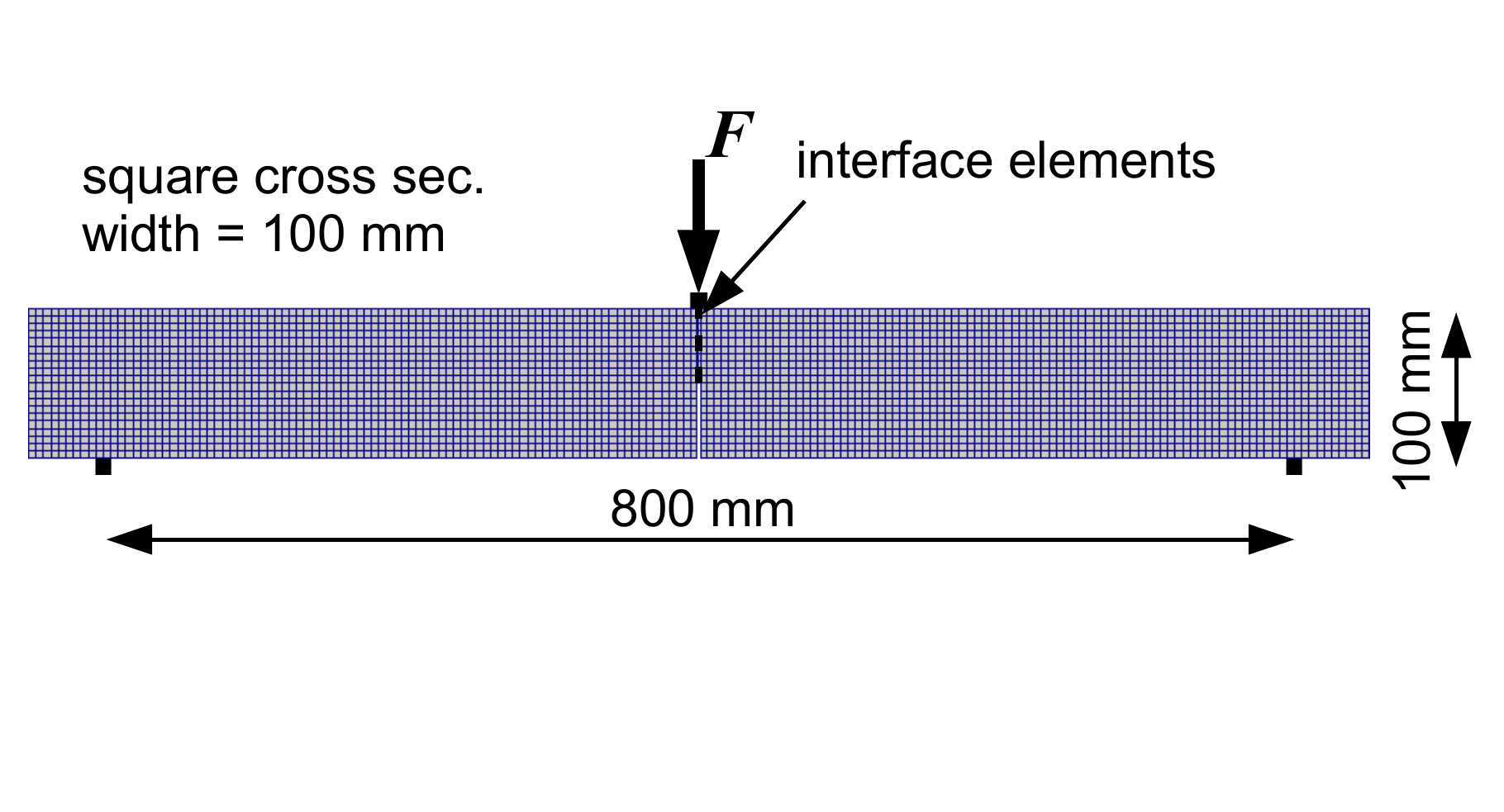} &
  \includegraphics[width=0.4\textwidth]{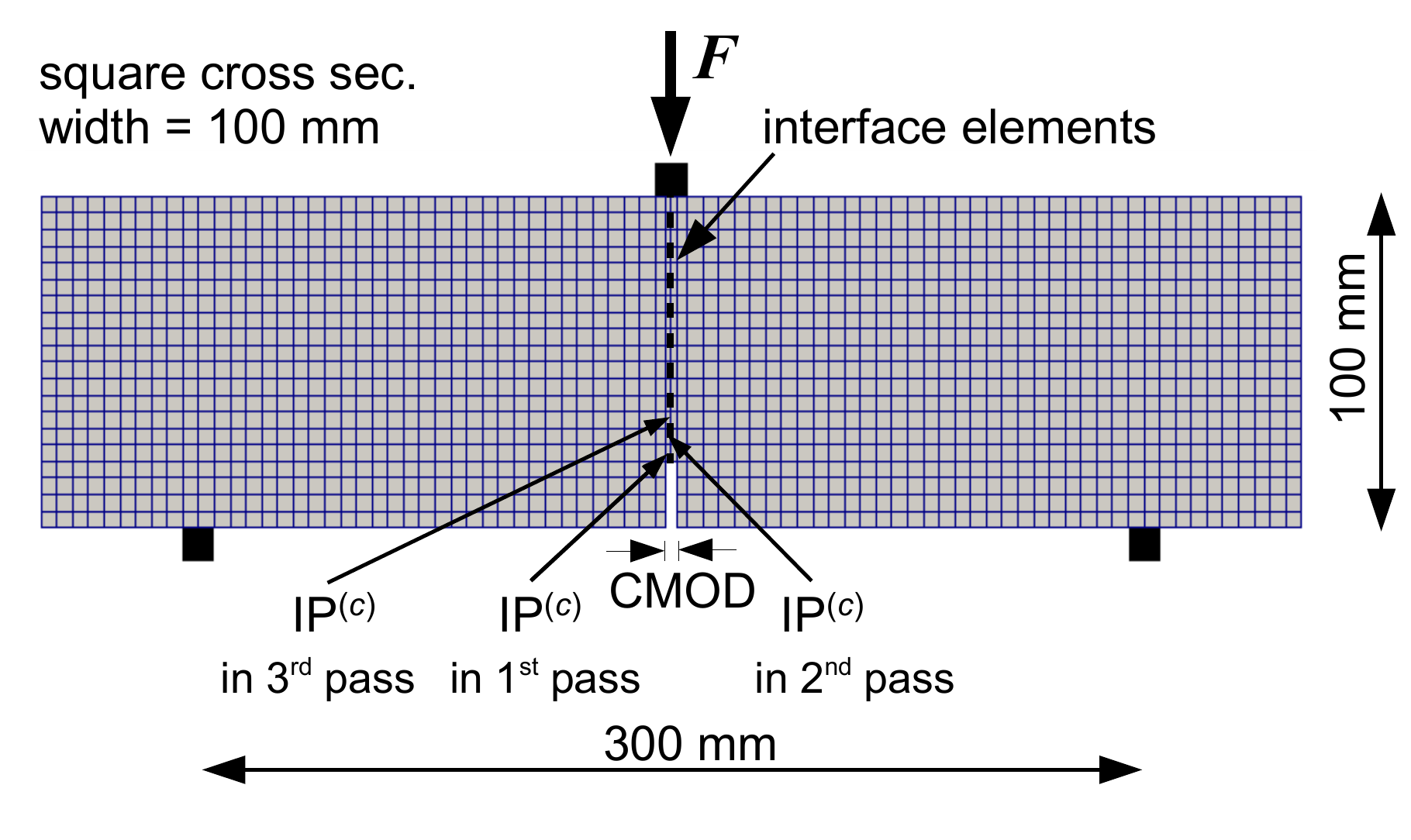} \\
  (a) & (b)
\end{tabular}
\caption{Setup of three point bending test presented by: a)~\protect\citet{Nanakorn:1996:Back}; b)~\protect\citet{Kabele:2015:modeling}}
\label{fig:3PB_setup}
\end{center}
\end{figure}

\paragraph{Fibre Reinforced cement-based composites} (FRCC) present a large group of materials with a variety of fracture properties. The reason for adding fibres to cementitious matrix is to overcome its brittleness by improving the post-cracking behaviour. Three-point bending tests are often used for the inverse analysis to determine the TS diagrams (\figref{fig:3PB_setup})~\cite{JCI:2003:japanconcrete,Nanakorn:1996:Back,Planas:2007:report39}. The crack is always assumed to propagate from the notch tip and to be straight. The necessary user defined parameters are summarised in~Tab.~\ref{tab:ECC_param}.

\begin{table}
\caption{Material properties of fibre reinforced cement-based composites \newtext{and user-defined parameters}} \label{tab:ECC_param}
\medskip\noindent\centering
\begin{tabular}{ccccc}
\hline Exp. No. & $E$ [GPa] & $\nu$ [-] & $k_0$ [MN/m] & $g_0$ [MN/m] \\
\hline
\multicolumn{5}{l}{\citet{Nanakorn:1996:Back}} \\
1  & 24.5 & 0.2  & 24.5 & 24.5$\cdot 10^4$ \\
\multicolumn{5}{l}{\citet{Kabele:2015:modeling}} \\
1  & 37.5 & 0.23 & 37.5 & 37.5$\cdot 10^4$ \\
2  & 37.0 & 0.23 & 37.0 & 37.0$\cdot 10^4$ \\
3  & 40.0 & 0.23 & 40.0 & 40.0$\cdot 10^4$ \\
\hline
\end{tabular}
\end{table}

\begin{figure}
\begin{center}
\begin{tabular}{cc}
  \includegraphics[width=0.4\textwidth]{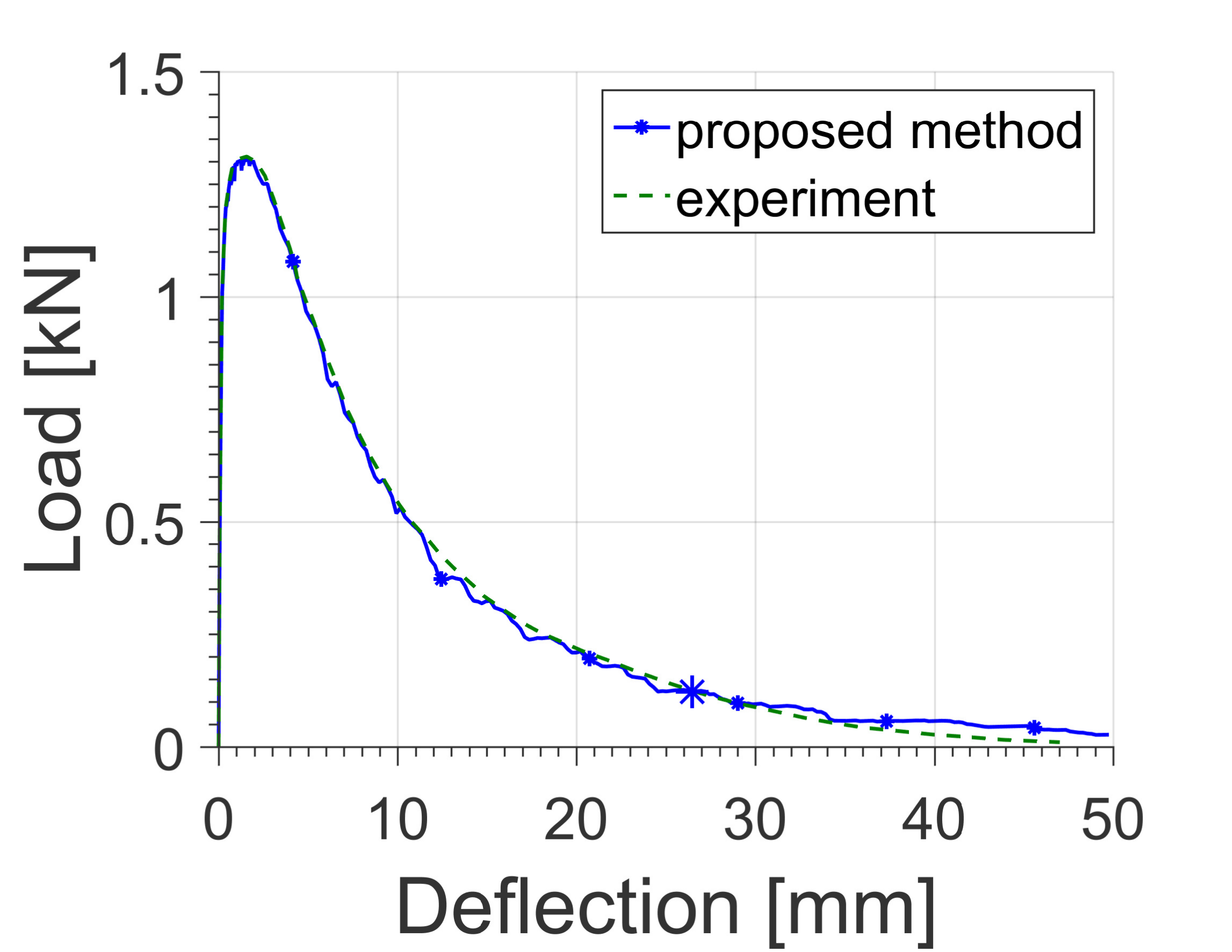} &
  \includegraphics[width=0.4\textwidth]{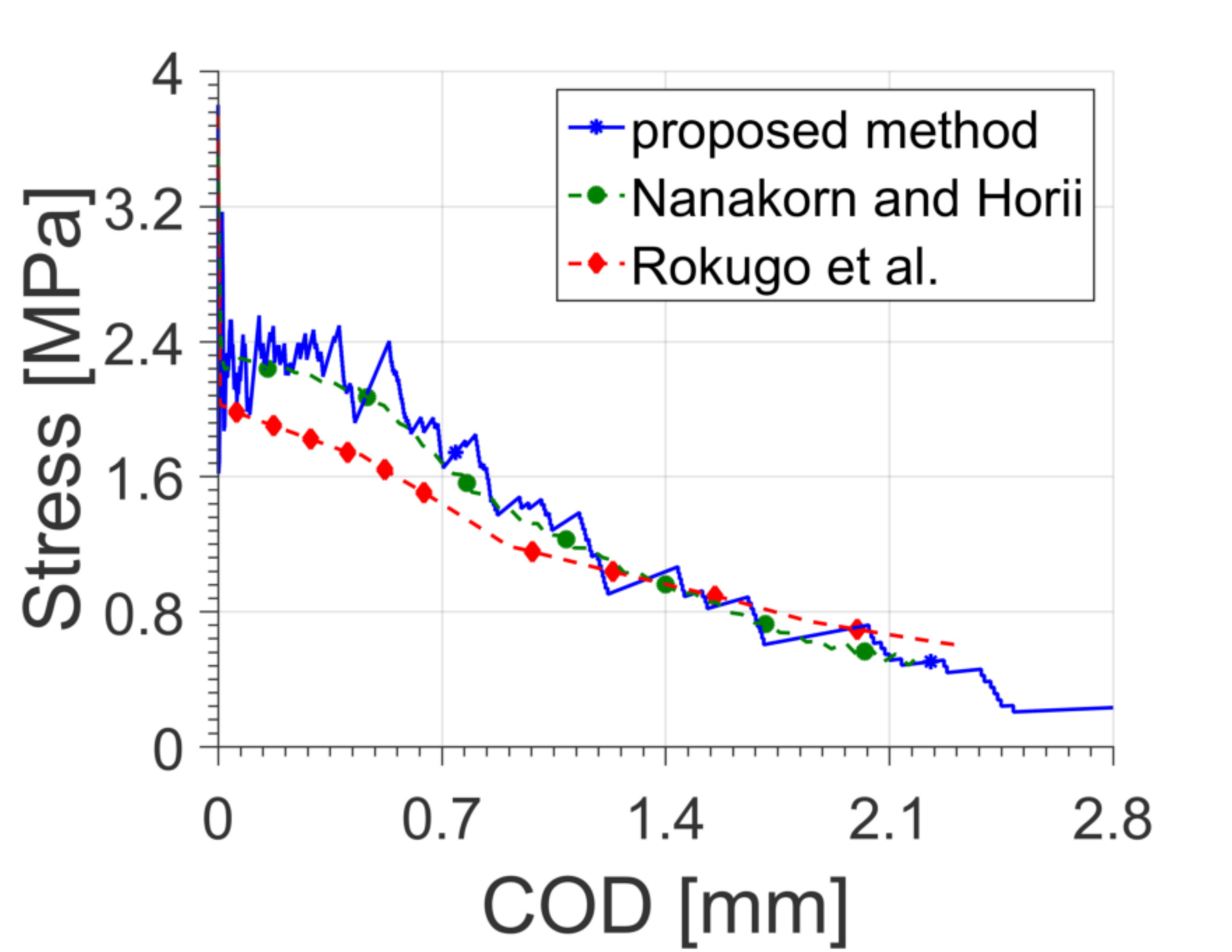} \\
  (a) & (b)
\end{tabular}
\caption{Three-point bending test used in~\protect\cite{Nanakorn:1996:Back}: a)~load-deflection diagrams; b)~TS diagrams}
\label{fig:3PB_Nanakorn}
\end{center}
\end{figure}

As the first example, we compare the performance of the proposed method with results reported by~\citet{Rokugo:1989:fracture} and~\citet{Nanakorn:1996:Back}. The load-deflection data from a three-point bending test (Figs.~\ref{fig:3PB_setup}(a) and \ref{fig:3PB_Nanakorn}(a))  from~\cite{Rokugo:1989:fracture} were used as input. The multi-pass enhancement was not used in this calculation. The calculated TS diagram is shown in~\figref{fig:3PB_Nanakorn}(b). The figure also contains the TS curve obtained by~\citet{Rokugo:1989:fracture} employing the J-integral method and that solved by inverse analysis by~\citet{Nanakorn:1996:Back}. It is obvious that the result of the proposed method shows some oscillations, which can be attributed to the inhomogeneity of fracture plane and discretisation of the input data. Similar oscillations were probably experienced also by~\citet{Nanakorn:1996:Back}, who eventually used a smoothing scheme to post-process the retrieved TS curve. However, the typical trends, i.e. initial stress drop and long tail, match very well the result in~\cite{Nanakorn:1996:Back}.

\begin{figure}
\begin{center}
\begin{tabular}{cc}
  \includegraphics[width=0.4\textwidth]{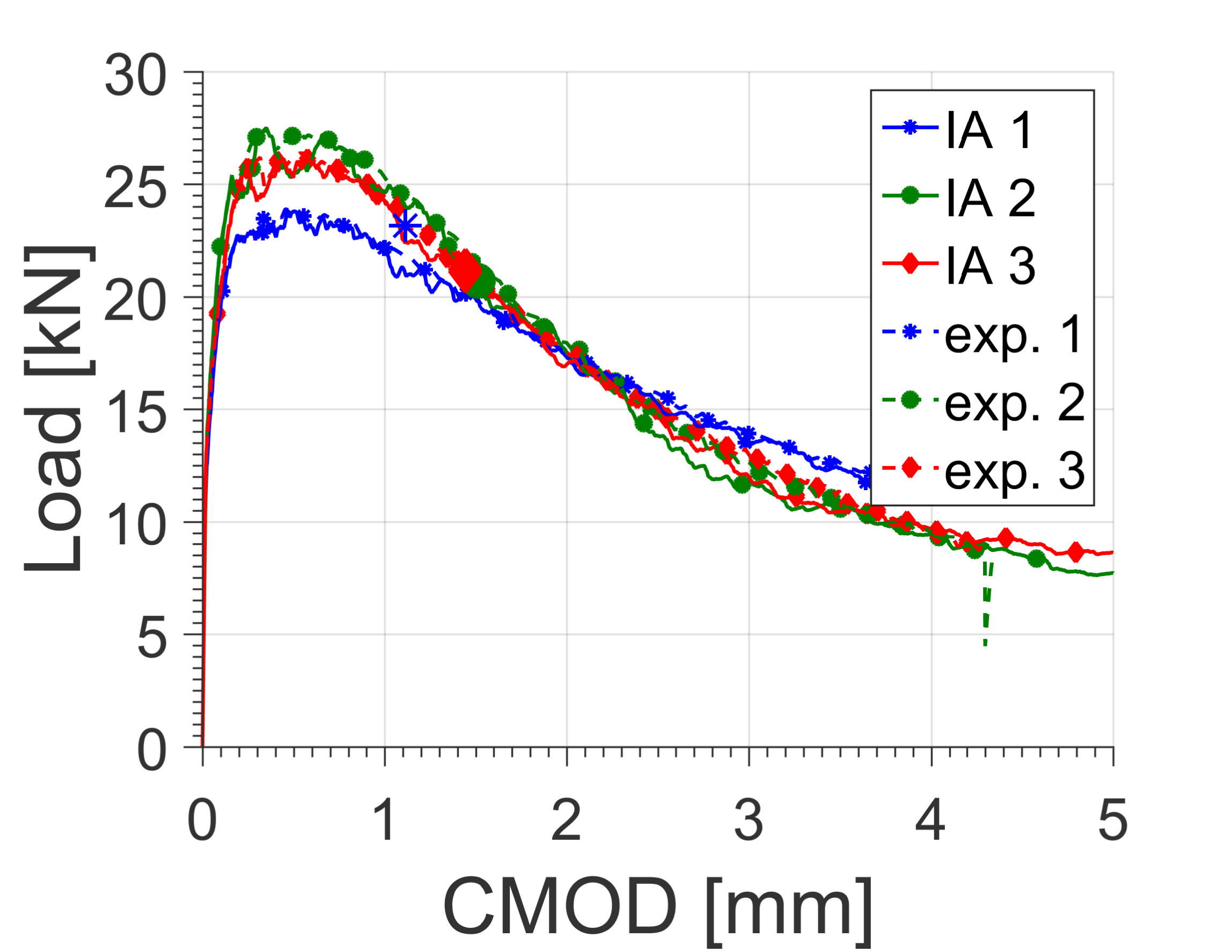} &
  \includegraphics[width=0.4\textwidth]{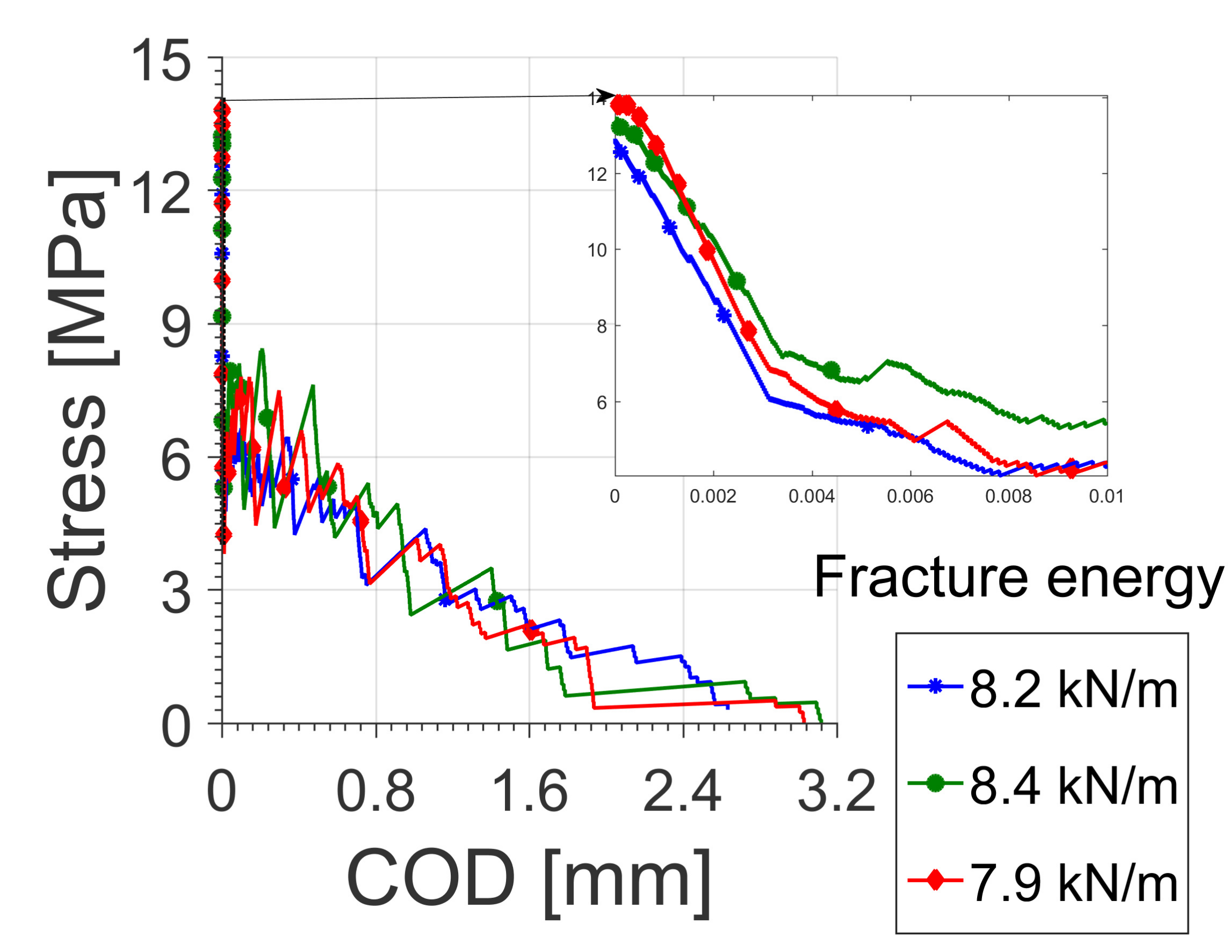} \\
  (a) & (b)
\end{tabular}
\caption{Three-point bending test of FRCC: a)~load-CMOD diagrams; b)~TS diagrams and corresponding fracture energies}
\label{fig:3PB_exp}
\end{center}
\end{figure}

As the second example, three sets of experimental data from three-point bending tests (\figref{fig:3PB_setup}(b)) on a high-strength fibre reinforced concrete~\cite{Kabele:2015:modeling} are used to determine TS diagrams, see~\figref{fig:3PB_exp}. The resulting TS diagrams in~\figref{fig:3PB_exp}(b) show good agreement if compared to each other in the sense of both shapes and fracture energies. As expected for this type of material, there is a rapid softening after the tensile strength is reached followed by a long tail~\cite{Lofgren:2005:fracture} and the calculated fracture energies correspond to results presented in~\cite{Nam:2017:experimental}.

\begin{figure}
\begin{center}
\begin{tabular}{cc}
  \includegraphics[width=0.4\textwidth]{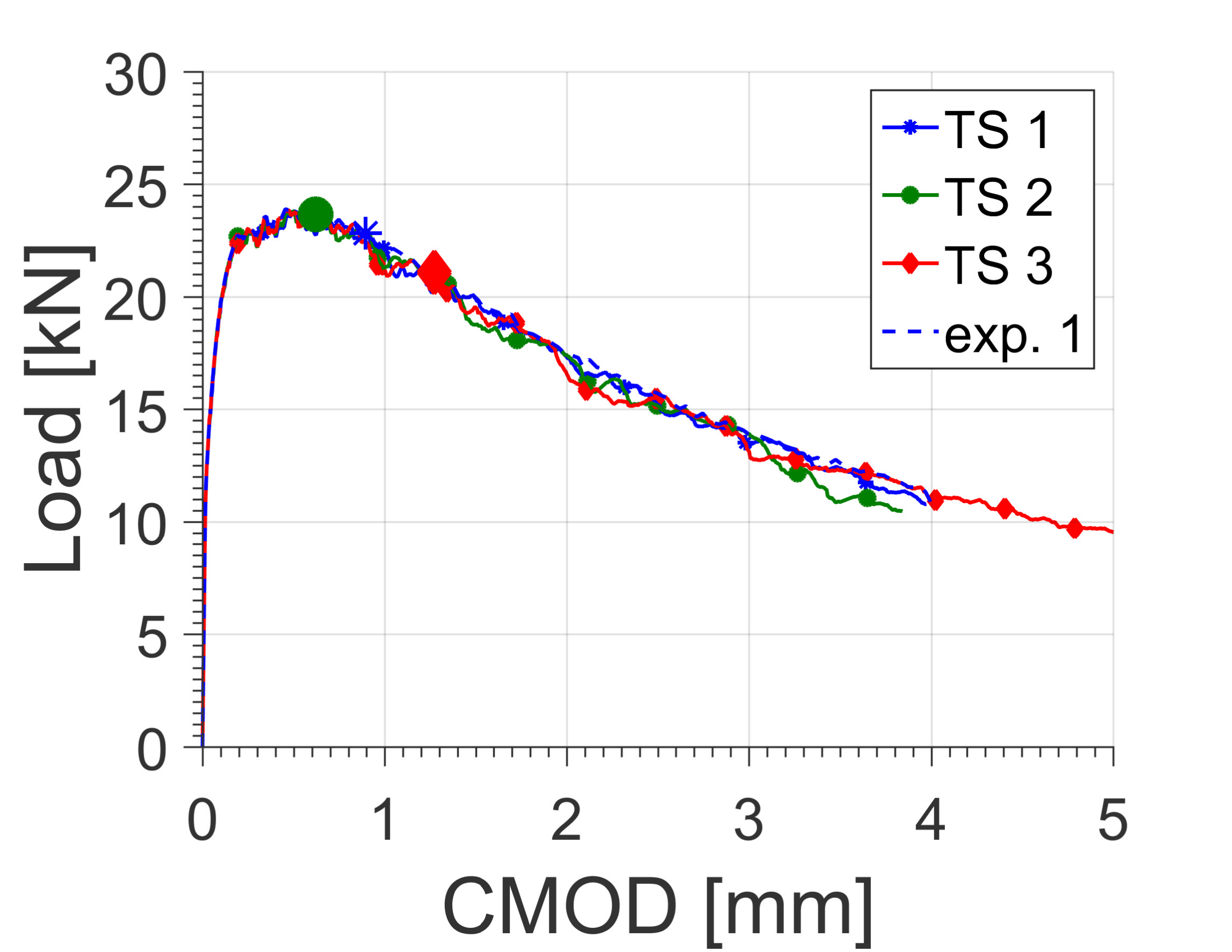} &
  \includegraphics[width=0.4\textwidth]{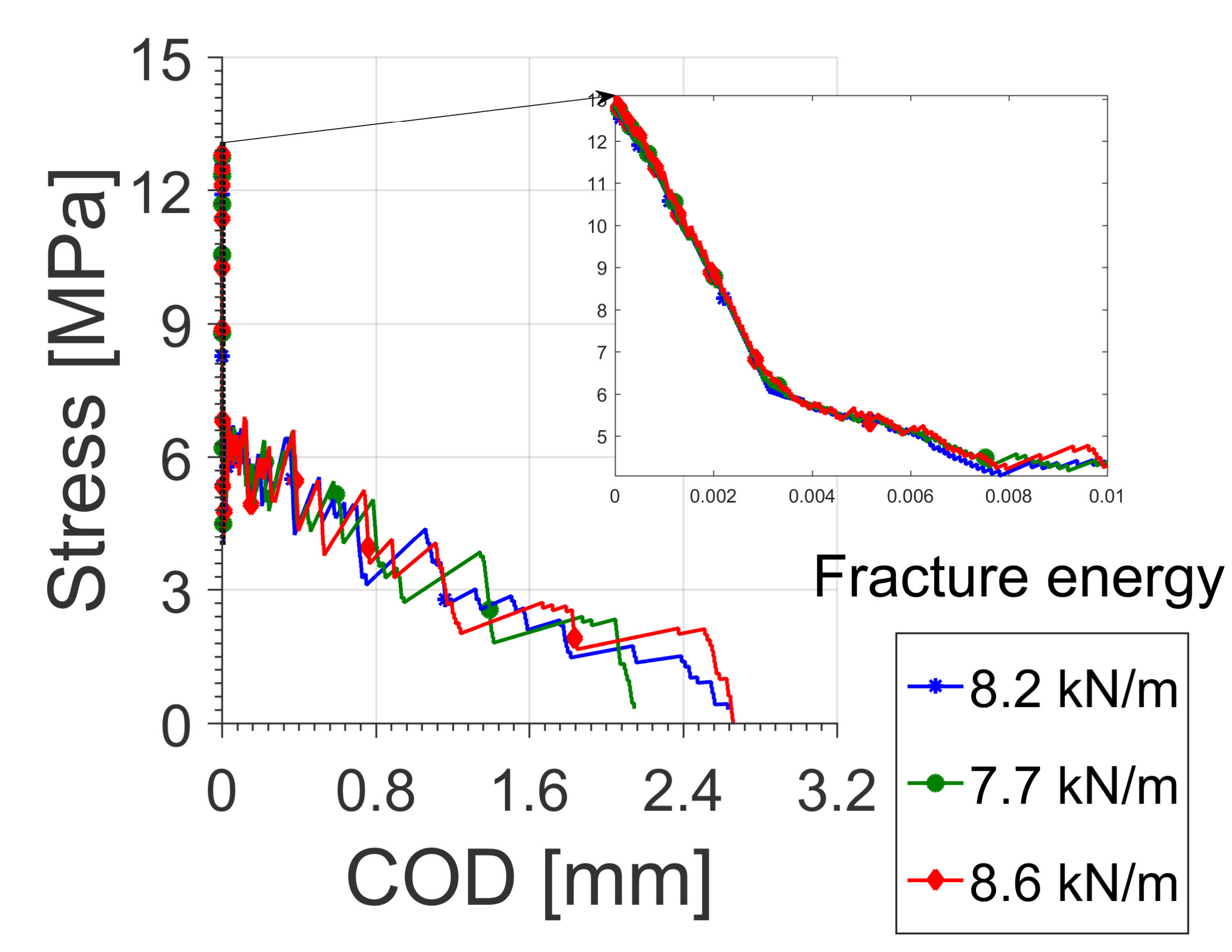} \\
  (a) & (b)
\end{tabular}
\caption{Three-point bending test of FRCC - multi-step approach: a)~load-CMOD diagrams; b)~TS diagrams and corresponding fracture energies calculated in individual passes of inverse analysis}
\label{fig:3PB_expMS}
\end{center}
\end{figure}

Additionally, the application of the multi-pass enhancement (see Section~\ref{subsec:multipass}), consisting of three subsequent passes, is demonstrated on the inverse analysis of the experiment No.~1. Let us recall that in each pass, the load-displacement (CMOD) data only up to the instant when the lead integration point becomes traction free, are used to determine the TS relation. These states are indicated by the large markers in~\figref{fig:3PB_expMS}(a). The TS relation is then assigned to the lead integration point, which was used to define it. In each subsequent pass, the lead integration point moves further along the crack path (\figref{fig:3PB_setup}(b)) and other TS relations are calculated. The obtained TS diagrams of each pass are shown in~\figref{fig:3PB_expMS}(b) (line types correspond to~\figref{fig:3PB_expMS}(a)). It is obvious, that the curves are fairly close, which indicates a good homogeneity of the ligament.

\begin{figure}
\begin{center}
\begin{tabular}{cc}
  \includegraphics[width=0.4\textwidth]{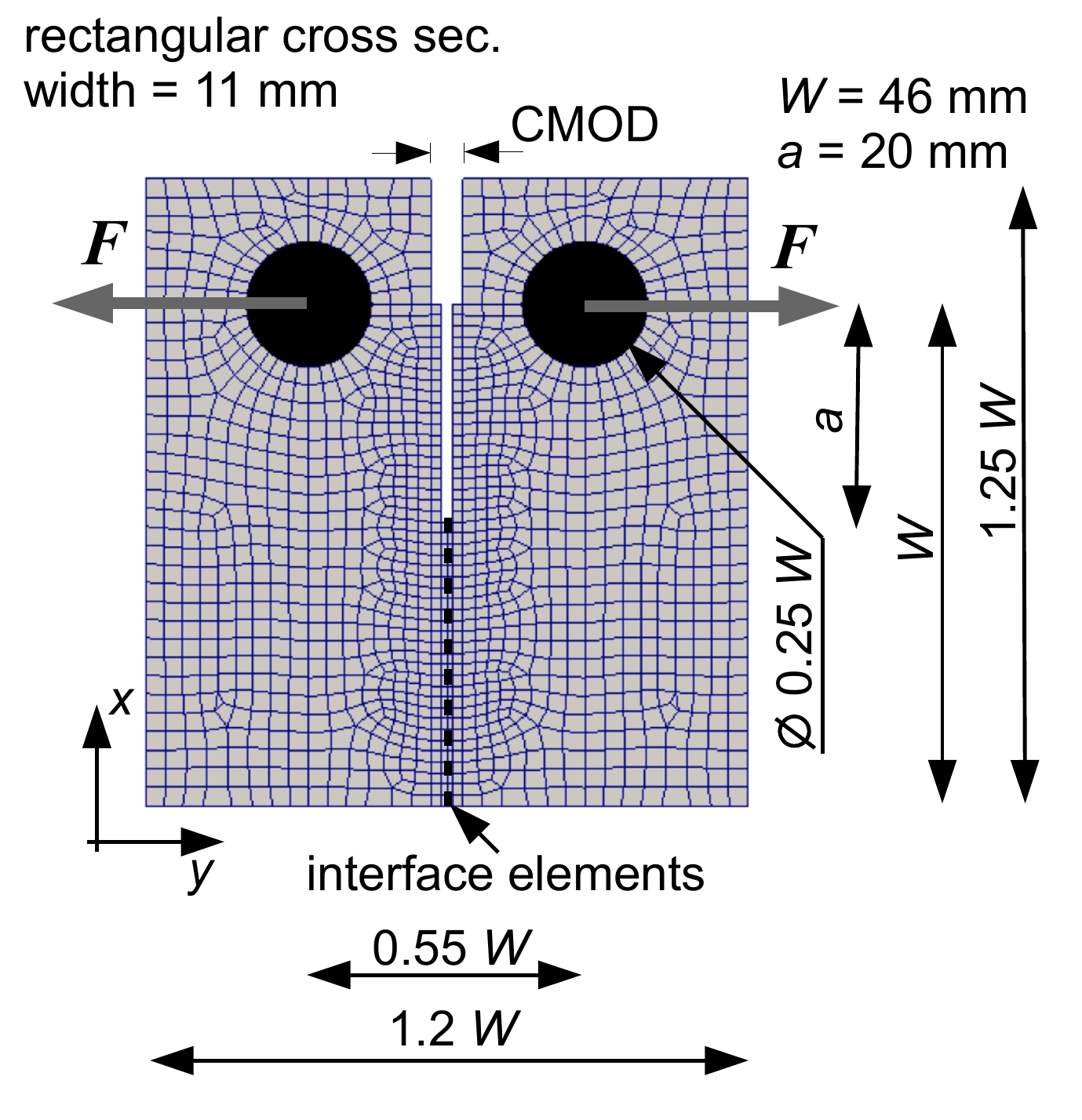}
\end{tabular}
\caption{Compact tension test setup}
\label{fig:split_setup}
\end{center}
\end{figure}

\paragraph{Wood} is an organic material which has been utilised in engineering practice for centuries. The experimental data of two compact tension tests (\figref{fig:split_setup}) presented in~\cite{Blass:2011:LWA} are used to obtain TS relations of wood. The tests were configured in such a way, that the initial notch as well as the propagating crack were parallel with the wood fibres.

In the present analysis, wood is modelled as a linear elastic orthotropic material with parameters listed in Tab.~\ref{tab:wood_param}. Values of Poisson's ratios are assumed based on~\cite{Kasal:2004:wood,Kasal:2013:wood}, while Young's and shear moduli are determined from the initial elastic part of the input load-CMOD curves adopted from~\cite{Blass:2011:LWA} and ranges presented in~\cite{Kasal:2004:wood,Kasal:2013:wood}. The element size in the region of interest is set to 1~mm.

\begin{figure}
\begin{center}
\begin{tabular}{cc}
  \includegraphics[width=0.4\textwidth]{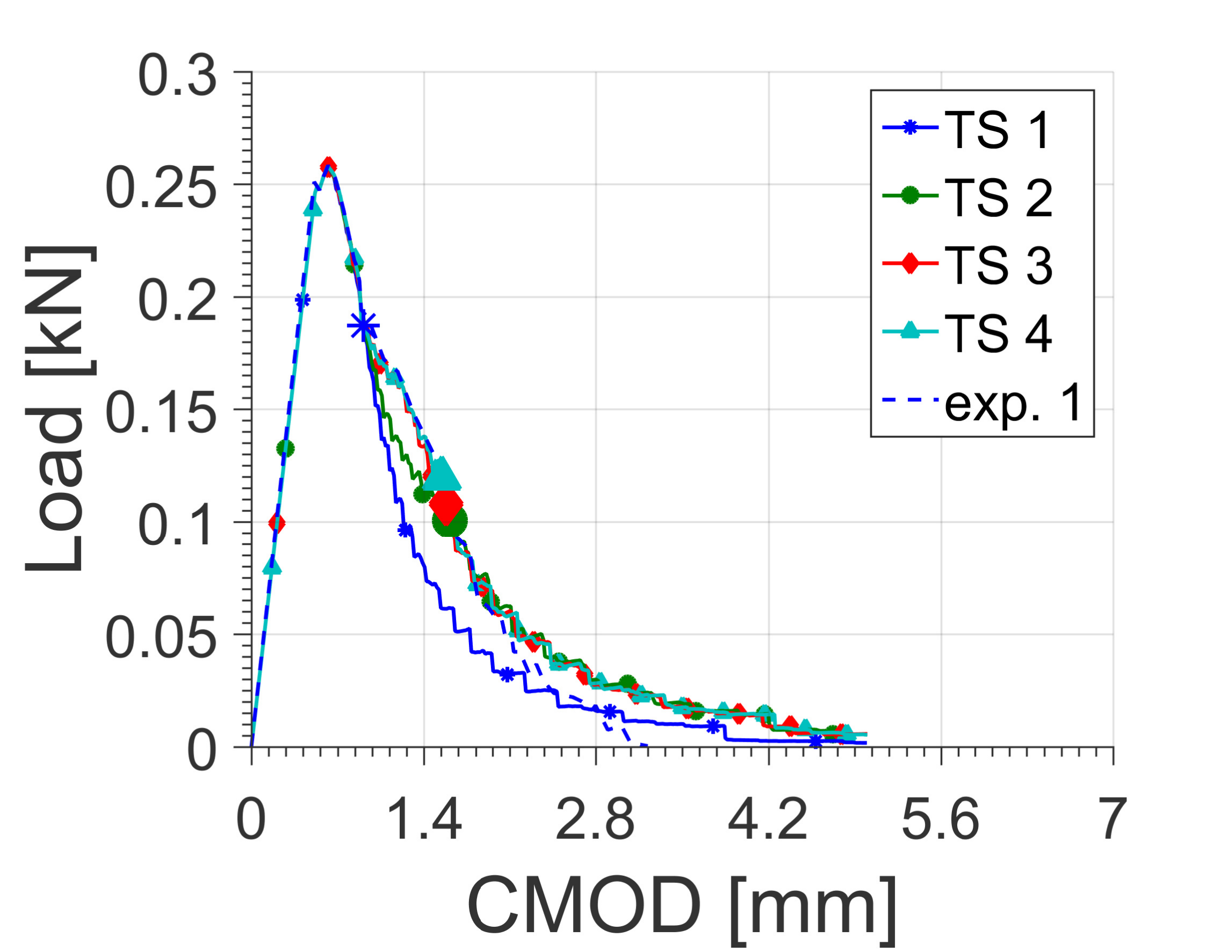} &
  \includegraphics[width=0.4\textwidth]{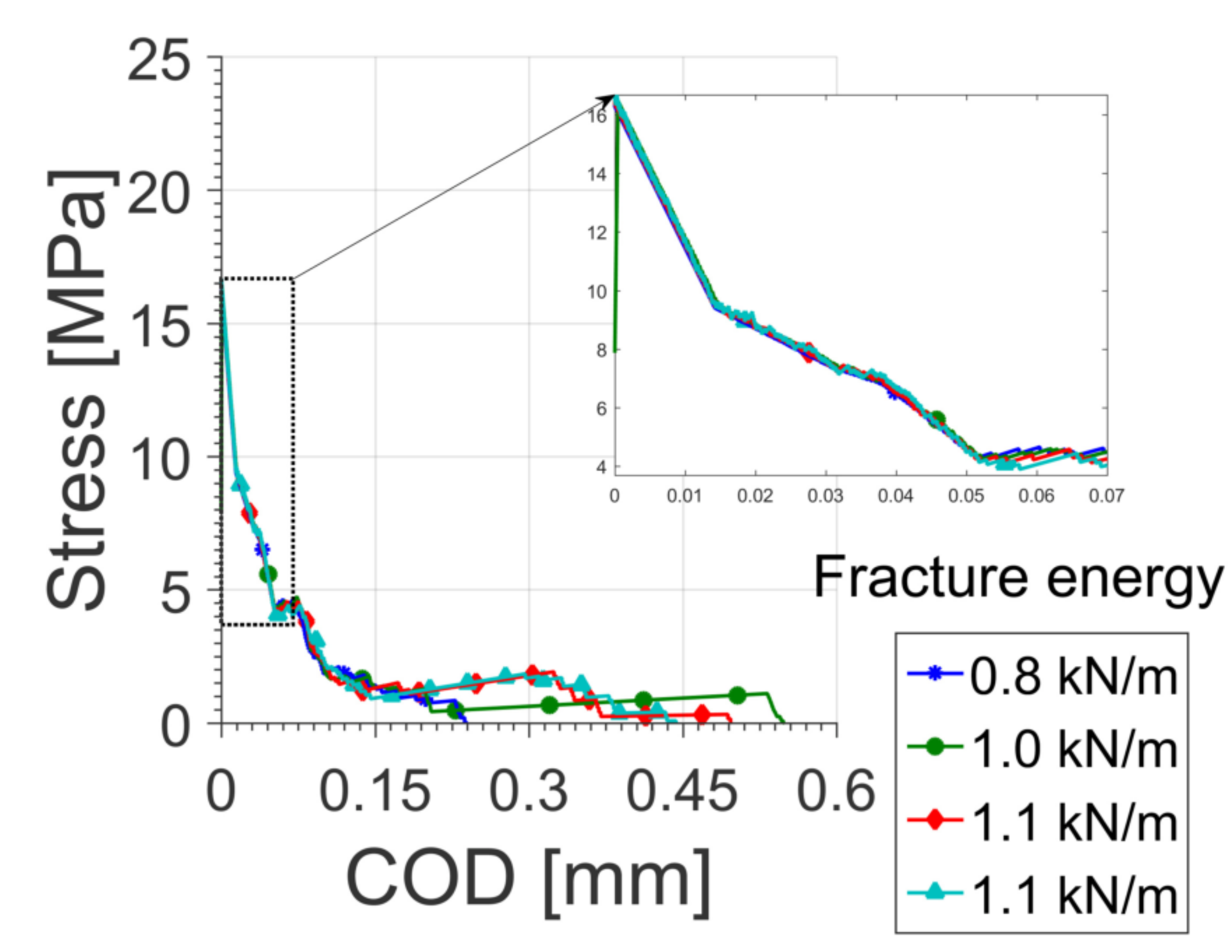} \\
  (a) & (b)
\end{tabular}
\caption{Compact tension test No.~1: a)~load-CMOD diagrams; b)~TS diagrams and corresponding fracture energies calculated in individual passes of inverse analysis}
\label{fig:splitno1}
\end{center}
\end{figure}

\begin{figure}
\begin{center}
\begin{tabular}{cc}
  \includegraphics[width=0.4\textwidth]{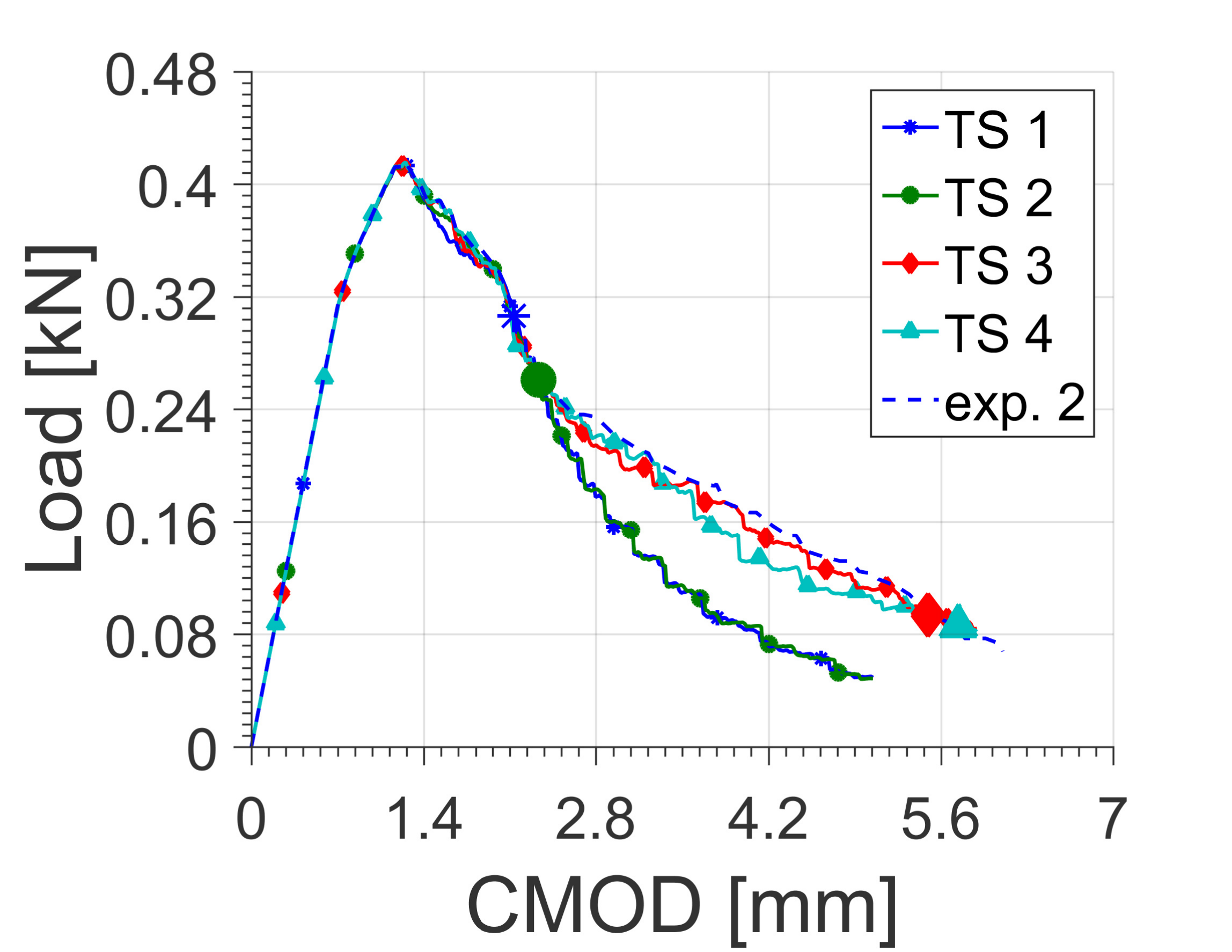} &
  \includegraphics[width=0.4\textwidth]{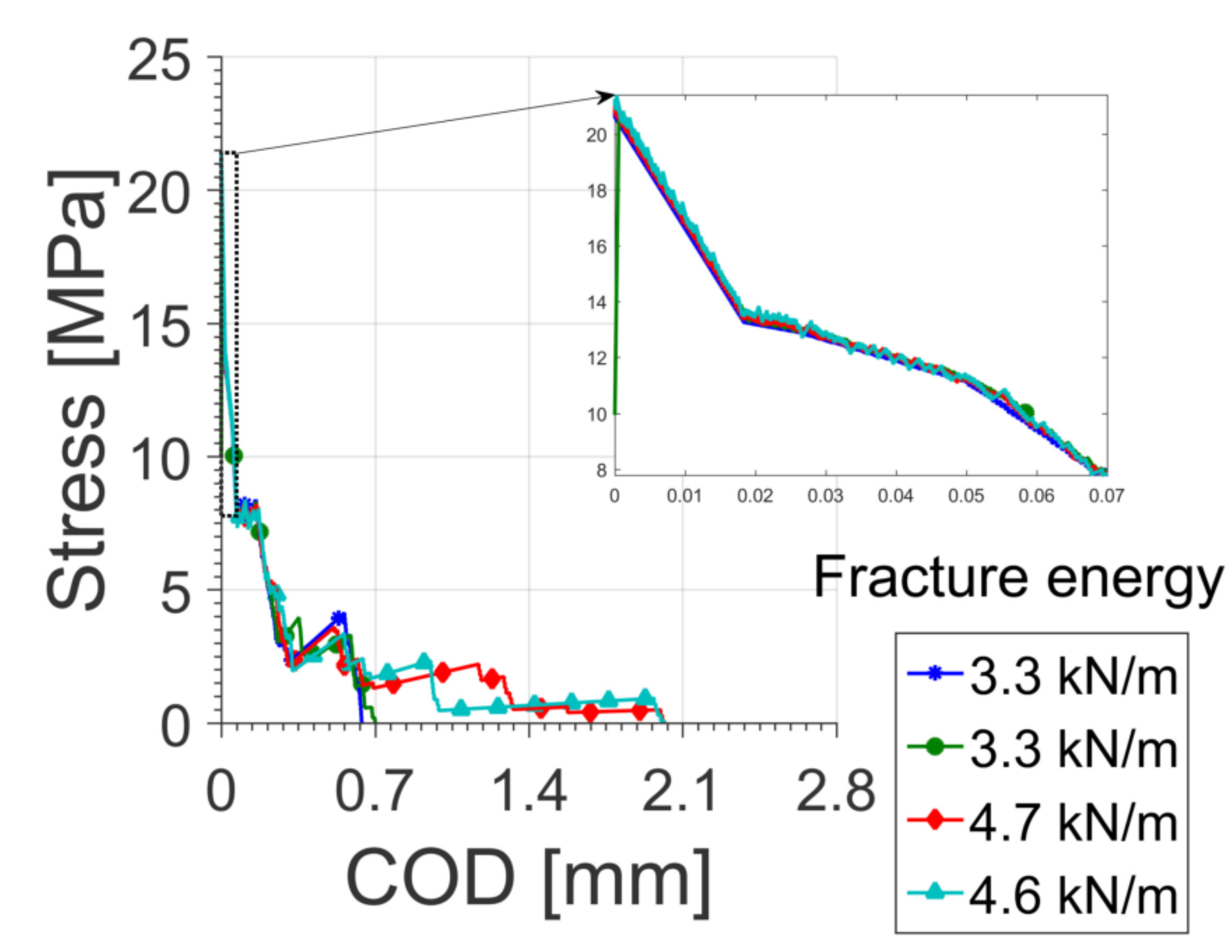} \\
  (a) & (b)
\end{tabular}
\caption{Compact tension test No.~2: a)~load-CMOD diagrams; b)~TS diagrams and corresponding fracture energies calculated in individual passes of inverse analysis}
\label{fig:splitno2}
\end{center}
\end{figure}

Figs.~\ref{fig:splitno1}(a) and \ref{fig:splitno2}(a) show the input load-CMOD curves (dashed lines) and the matched curves from individual passes of the inverse analysis (solid lines). The TS curves from each pass are depicted in~Figs.~\ref{fig:splitno1}(b) and \ref{fig:splitno2}(b). In this example, we continued the calculation in each pass even beyond the state when the lead integration point became traction free and the TS relation was fully identified. Thus, the portions of the load-CMOD curves in~Figs.~\ref{fig:splitno1}(a) and \ref{fig:splitno2}(a) beyond the large markers show the predicted behaviour of the specimen when the TS relations from the respective passes are used (e.g. in pass 1, all cracked integration points use TS~1, in pass 2, the first lead integration point uses TS~1 and remaining integration points use TS~2, etc). By comparing the solid lines with the dashed experimental curve beyond the big markers it is obvious, that if the TS relation is based only on the first lead integration point, the prediction considerably deviates from the experimental result. By multiple passes, the accuracy of the match between the experimental curve and the simulation is improved. This finding is consistent with the difference among the individual TS curves in~Figs.~\ref{fig:splitno1}(b) and \ref{fig:splitno2}(b), where it is obvious that the curves mostly differ at larger crack opening displacements. This indicates that the cohesive property along the crack path has a larger variability, which could be expected in case of a natural material, such as wood.

\begin{table}
\caption{Material properties of wood \newtext{and user-defined parameters}} \label{tab:wood_param}
\medskip\noindent\centering\small
\begin{tabular}{cccccccc}
\hline %
Exp. No. & $E_{xx}$ & $E_{yy},E_{zz}$ & $G_{xy},G_{xz}$ & $G_{yz}$ & $\nu_{yz},\nu_{xz},\nu_{xy}$ & $k_0$ & $g_0$ \\
& [GPa] & [GPa] & [GPa] & [GPa] & [-] & [MN/m] & [MN/m] \\
\hline
1  & 18.90 & 0.80 & 0.62 & 0.19 & 0.3 & 18.9 & 18.9$\cdot 10^4$ \\
2  & 18.90 & 0.74 & 0.57 & 0.19 & 0.3 & 18.9 & 18.9$\cdot 10^4$ \\
\hline
\end{tabular}
\end{table}

\newtext{In summary, the proposed method was utilised to analyse real experimental data for a fibre reinforced cement-based composites and wood, see also~\cite{Czernuschka:2018:SLA} for additional examples. As can be seen from the presented results, the procedure produced reliable traction-separation diagrams, especially if the multi-pass approach was employed.}

\section{Conclusions}
\label{sec:concl}
\oldtext{A new procedure for the inverse identification of traction-separation relationship based on loading curve from a fracture test was proposed in the present paper. First, the method was verified against input data obtained from numerical ``artificial'' experiments, for which the traction-separation diagram was known. As can be seen from the presented results of sensitivity study, the procedure provides reasonably good results even for the user-defined parameters, which are set with a relatively big error compared to the true values. Secondly, the method was utilised to analyse real experimental data for a fibre reinforced cement-based composites and wood. The procedure produced consistent traction-separation diagrams, especially if the multi-pass approach was employed. The multi-pass approach can capture the variability of cohesive fracture properties along the crack path (Figs.~\ref{fig:splitno1} and \ref{fig:splitno2}) and can also improve the agreement between the experimental data and the numerical analysis.}

\newtext{A traction-separation relationship is an important material characteristic, which is commonly used to describe the complex physical phenomena underlying the initiation and evolution of damage in the process zone of quasi-brittle solids~\cite{Hoover:2013:Test,Cusatis:2009:cohesive,Grassl:2012:Meso}. A new procedure for the inverse identification of traction-separation relationship based on loading curve from a fracture test was proposed in the present paper.}

In general, the proposed procedure has the following advantages:
\newtext{
\begin{itemize}
  \item 
      Applicability with a variety of experimental data - the proposed procedure provides outstanding versatility for evaluation of TS diagrams from a variety of experimental configurations and measurements characterized by various loading vs. response data pairs, such as, load-deflection (displacement), load-CMOD.
  \item 
        Limited number of input parameters - in contrast to other methods~\cite{Nanakorn:1996:Back,Kitsutaka:1995:fracture,Su:2012:Incremental}, neither tensile strength nor fracture energy have to be specified as an input for the proposed method.
  \item Generality of the identified TS relation - in a difference with other methods~\cite{Hordijk:1991:local,Skocek:2008:Inverse}, the proposed method identifies individual points of the TS relationship, which does not have to follow any pre-determined functional form. The TS relation may, for example, involve multiple hardening and softening intervals.

  \item Appealing numerical stability - the procedure is based on a sequence of linear (secant) steps followed by a stiffness reduction which ensures a good numerical stability of the calculations.
  \item Solution uniqueness - the identified TS relation is unique for a given set of input data, parameters and FE discretization since the incremental inverse analysis, in which each point of the TS relation is found in a direct way, is used. Note that by performing several passes over the input data (Section~\ref{subsec:multipass}), multiple TS relations can be found, as each time, a different portion of the input load-displacement curve and a different lead IP are utilised.
  \item 
       Compatibility with various constitutive models - a variety of constitutive models compatible with the sequentially linear approach can be employed to characterize the material outside the cracking zone.
\end{itemize}
}

\section*{Acknowledgements}
The first author would like to gratefully acknowledge the financial support provided by the GA\v{C}R grant No.~15-10354S. The second author would like to acknowledge the support by the GA\v{C}R grant No.~13-15175S.

\bibliographystyle{elsarticle-num-names}
\bibliography{liter}

\end{document}